\documentclass[twocolumn,english,superscriptaddress,amsmath,amssymb,floats,prb,longbibliography]{revtex4-2}
\usepackage[latin9]{inputenc}
\usepackage{color}
\usepackage{mathtools}
\usepackage{amsmath}
\usepackage{graphicx}
\usepackage{multirow,booktabs}
\usepackage{array}
\usepackage{xcolor,colortbl}
\usepackage[normalem]{ulem}
\usepackage{lipsum}
\usepackage{soul}

\newcommand{\mbf}[1]{\mathbf{#1}}

\makeatletter

\usepackage{comment}

\newcommand{\AVS}{$A$V$_3$Sb$_5$}

\newcommand{\QQ}{$3\mathbf{Q}$-$3\mathbf{Q}$}
\newcommand{\Q}{$2\mathbf{Q}$-$1\mathbf{Q}$}

\@ifundefined{textcolor}{}{%
 \definecolor{BLACK}{gray}{0}
 \definecolor{WHITE}{gray}{1}
 \definecolor{RED}{rgb}{1,0,0}
 \definecolor{GREEN}{rgb}{0,1,0}
 \definecolor{BLUE}{rgb}{0,0,1}
 \definecolor{CYAN}{cmyk}{1,0,0,0}
 \definecolor{MAGENTA}{cmyk}{0,1,0,0}
 \definecolor{YELLOW}{cmyk}{0,0,1,0}
}

\usepackage{epstopdf}
\usepackage{array}
\usepackage{mathrsfs}
\usepackage{dcolumn}
\usepackage{bm}

\usepackage{xcolor}

\newcolumntype{M}[1]{>{\centering\arraybackslash}m{#1}}
\newcolumntype{P}[1]{>{\centering\arraybackslash}p{#1}}

\newcolumntype{L}[1]{>{\raggedright\let\newline\\\arraybackslash\hspace{0pt}}m{#1}}
\newcolumntype{C}[1]{>{\centering\let\newline\\\arraybackslash\hspace{0pt}}m{#1}}
\newcolumntype{R}[1]{>{\raggedleft\let\newline\\\arraybackslash\hspace{0pt}}m{#1}}

\graphicspath{{../Inkscape/},{../Mathematica/},{./Figures/}}

\usepackage{tikz}
\usetikzlibrary{calc,intersections}
\usetikzlibrary{shadings}
\usepgflibrary{shadings}

\usepackage{babel}

\begin{document}

\title{Loop currents in $A$V$_3$Sb$_5$ kagome metals: multipolar and toroidal magnetic orders}

\author{Morten H. Christensen}
\email{mchriste@nbi.ku.dk}
\affiliation{Niels Bohr Institute, University of Copenhagen, 2100 Copenhagen, Denmark}

\author{Turan Birol}
\affiliation{Department of Chemical Engineering and Materials Science, University of Minnesota, MN 55455, USA}

\author{Brian M. Andersen}
\affiliation{Niels Bohr Institute, University of Copenhagen, 2100 Copenhagen, Denmark}

\author{Rafael M. Fernandes}
\affiliation{School of Physics and Astronomy, University of Minnesota, Minneapolis,
MN 55455, USA}

\date{\today}
\begin{abstract}
Experiments in the recently discovered vanadium-based kagome metals have suggested that their charge-ordered state displays not only bond distortions, characteristic of a ``real" charge density-wave (rCDW), but also time-reversal symmetry-breaking, typical of loop currents described by an ``imaginary" charge density-wave (iCDW). Here, we combine density-functional theory, group-theory, and phenomenological modeling to investigate the complex charge-ordered states that arise from interactions between the low-energy van Hove singularities present in the electronic structure of \AVS. We find two broad classes of mixed iCDW-rCDW configurations: triple-$\mathbf{Q}$ iCDW, triple-$\mathbf{Q}$ rCDW order, dubbed \QQ{}, and double-$\mathbf{Q}$ iCDW, single-$\mathbf{Q}$ rCDW order, dubbed \Q{}. Moreover, we identify seven different types of iCDW order, stemming from the different vanadium-orbital and kagome-sublattice structures of the two pairs of van Hove singularities present above and below the Fermi level. While the \Q{} states trigger an orthorhombic distortion that breaks the threefold rotational symmetry of the kagome lattice, the \QQ{} states induce various types of subsidiary uniform magnetic orders, from conventional ferromagnetism to magnetic octupolar, magnetic toroidal, and even magnetic monopolar order. We show that these exotic orders display unique magneto-striction, magneto-electric, and magneto-electric-striction properties that can be probed experimentally to identify which iCDW state is realized in these compounds. We briefly discuss the impact of an out-of-plane modulation of the charge order and the interplay between these complex charge-ordered states and superconductivity.
\end{abstract}
\maketitle

\section{Introduction} 

Systems in which electronic correlations and nontrivial topology are simultaneously present are of major interest for the condensed matter community. The kagome lattice~\cite{Syozi1951} offers a promising platform to realize these phenomena~\cite{Ghimire2020Topology}, as its electronic structure exhibits Dirac cones~\cite{Guo2009,Ye2018,Kang2020}, flat bands~\cite{Lin2018,Meier2020}, and van Hove singularities~\cite{Kang2022Twofold}. Hence, the recent discovery of a new family of superconducting (SC) kagome materials~\cite{Ortiz2019New}, \AVS{} ($A$=K, Rb, Cs), with $T_c \sim 2$~K depending on the alkali atom~\cite{Ortiz2020CsV3Sb5,Ortiz2021Superconductivity,Yin2021Superconductivity}, was met with great enthusiasm by the community~\cite{Ghimire2020Topology,Jiang2021Kagome,Christensen2022Electrons,Neupert2022Charge}. The nature of the SC state remains widely debated: while experiments have reported both nodeless ~\cite{Mu_2021,Duan2021,Xu_PRL_2021,Chen2021Roton,Roppongi2022Bulk} and nodal behavior~\cite{Guguchia2022Tunable,Zhao_2021}, theoretical models have proposed unconventional chiral $d$-wave and $f$-wave states~\cite{Kiesel2013Unconventional,Wang2013Competing,Wu2021Nature,Wen2022}. More exotic SC phenomena have also been discussed, motivated by intriguing data, including a time-reversal symmetry-breaking pairing state \cite{Mielke2022Time-reversal,Guguchia2022Tunable,Khasanov2022Charge,Gupta2022}, a pair density-wave~\cite{Chen2021Roton}, and charge-$4e$ and charge-$6e$ condensates~\cite{Ge2022Discovery}. 

Superconductivity in these materials, however, emerges inside a charge density-wave (CDW) state, which onsets at $T_{\rm CDW}\sim 100$~K~\cite{Ortiz2019New,Stahl2021Temperature}. Importantly, $T_{\rm CDW}$ and $T_c$ anti-correlate as a function of pressure, uniaxial stress, and doping~\cite{Yu2021Unusual,Chen2021Double,Qian2021Revealing,Oey2022Fermi}. Therefore, to achieve a complete description of the SC phase, much of the experimental and theoretical research has focused on elucidating the properties of the charge-ordered state~\cite{Yang2020Giant,Jiang2021Unconventional,Xiang2021,Chen2021Roton,Ratcliff2021,Uykur2022Optical,Ortiz2021Fermi,Li2021Spectroscopic,Lou2022Charge-Density-Wave-Induced,Kang2022Twofold,Li2021Observation,Luo2022Electronic,Mielke2022Time-reversal,Li2022Rotation,Nie2022Charge,Wu2022Charge,Khasanov2022Charge,Yu2021Evidence, Park2021,Lin2021,Denner2021,Tan2021,Christensen2021,Feng2021,Setty2021Electron,Feng2021Low-energy,ZWang2021,Tazai2022,PRM_Subedi2022}. It is well-established that the CDW leads to a $2 \times 2$ increase of the unit cell in the $(a,b)$ plane~\cite{Jiang2021Unconventional}, whereas along the crystallographic $c$-axis the increase can be a factor of $1$, $2$ or $4$ depending on the alkali atom, pressure, and temperature \cite{Ortiz2021Fermi,Stahl2021Temperature,Wu2022Charge,Li2021Observation,Miao2022}. The threefold rotational symmetry of the lattice has also been reported to be broken at either $T_{\rm CDW}$ or well below the onset of CDW order ~\cite{Xiang2021,Zhao2021Cascade,Li2022Rotation,Nie2022Charge,Xu_MOKE}. Consistent with these observations, a second CDW phase has been observed to emerge in the phase diagram of certain compounds \cite{Gupta2022,Miao2022}. These results are compatible with a scenario in which multiple CDW states with wave-vectors sharing the same in-plane components $(1/2, 1/2, Q_z)$ have comparable energy scales, as suggested by first-principles calculations~\cite{Tan2021,Ratcliff2021,Christensen2021,PRM_Subedi2022}.

The most surprising property of the CDW state, which makes it stand out compared to other CDW phases like those seen in metallic chalcogenides \cite{DiSalvo1975,DiSalvo1979,Rossnagel2011,Huang2020}, is that it appears to break time-reversal symmetry, as reported by scanning tunneling microscopy (STM), muon spin resonance ($\mu$SR), and magneto-optical Kerr effect (MOKE) measurements~\cite{Jiang2021Unconventional,Mielke2022Time-reversal,Guguchia2022Tunable,Khasanov2022Charge,Yu2021Evidence,Xu_MOKE,Hu2022Time-reversal}. A time-reversal symmetry-broken charge order has a natural interpretation in terms of periodic patterns of loop currents~\cite{Varma1997Non-Fermi-liquid,Nayak2000Density-wave}, reminiscent of the Haldane model for the quantum Hall effect~\cite{Haldane1988Model}. Theoretically, these loop-currents are described in terms of an ``imaginary'' CDW (iCDW) order parameter -- to be contrasted with a ``real'' CDW (rCDW) order parameter describing bond distortions or charge variations at the lattice sites~\cite{Nayak2000Density-wave,Kang2011,Chubukov2015}. The existence of such loop-currents is further supported by recent measurements of field-tuned chiral transport in CsV$_3$Sb$_5$~\cite{Guo2022Field-tuned}. Previous works have shown that the combination of van Hove singularities (vHs) in the electronic dispersion and electron-electron interactions can promote such iCDW states \cite{Park2021,Lin2021,Denner2021}. Importantly, the symmetries of the kagome lattice generally entangle the iCDW and rCDW order parameters, creating complex charge-order patterns that display both loop-currents and bond distortions~\cite{Park2021,Lin2021,Feng2021Low-energy}. Moreover, while the symmetry of the rCDW order parameter can be inferred from first-principles and recent phonon spectroscopy data~\cite{Ratcliff2021,Tan2021,Christensen2021,PRM_Subedi2022,Wu2022Charge}, the symmetry of the iCDW state remains unsettled. Thus, distinguishing the unique signatures of the many allowed mixed iCDW-rCDW configurations is paramount to establish the origin of the charge-order instability and, ultimately, the properties of the normal state from which superconductivity emerges.

In this paper, we combine group theory, first-principles calculations, and phenomenology to classify the viable iCDW-rCDW configurations that arise from interactions between the low-energy electronic states associated with the vHs of \AVS{}, which are located near the three symmetry-equivalent M points of the Brillouin zone (BZ). The key point is that there are multiple low-energy vHs -- a pair of points above and a pair of points below the Fermi energy -- with different vanadium orbital characters ($d_{z^2}$, $d_{xz}$, and $d_{yz}$) and distinct vanadium sublattice polarization ($p$-type, corresponding to ``pure" sublattice polarization, and $m$-type, corresponding to ``mixed" sublattice polarization~\cite{Kiesel2012,Denner2021}). From this rich landscape, we find various intra-orbital and inter-orbital loop-current patterns, resulting in seven different types of iCDW order parameters, corresponding to seven different irreducible representations (irreps) of the space group. In the presence of spin-orbit coupling, an iCDW transition necessarily triggers a spin density-wave (SDW) at the same wave-vectors \cite{Klug2018}, which could in principle be probed by neutron scattering. By using group theory, we show that the most natural way to understand the SDW is in terms of magnetic moments on the in-plane Sb ions, whose directions (i.e., in-plane versus out-of-plane) depend on the orbital character of the iCDW order. 

Remarkably, all cases considered here display the same iCDW-rCDW coupled Landau free-energy. By analysing the minima of this free energy, we find two general types of mixed iCDW-rCDW configurations for each of the seven iCDW order parameters. The first one is a $3\mathbf{Q}$-$3\mathbf{Q}$ configuration, in which the iCDW and rCDW order parameters both condense at the three distinct M wave-vectors. The second one is a $2\mathbf{Q}$-$1\mathbf{Q}$ configuration, where the iCDW order parameter condenses at two distinct M wave-vectors and the rCDW order parameter condenses at the remaining M wave-vector. We note that, while the first type of mixed iCDW-rCDW configuration was discussed in Ref.~\onlinecite{Park2021} for the case of intra-orbital iCDW order, here we also consider inter-orbital iCDW states as well as the $2\mathbf{Q}$-$1\mathbf{Q}$ configuration.

Given the challenges in directly probing loop-currents experimentally, we also analyze the experimental manifestations of these different mixed iCDW-rCDW configurations. Besides the finite-$q$ magnetism (i.e., SDW) triggered by the iCDW alone, the mixed iCDW-rCDW phases with $3\mathbf{Q}$-$3\mathbf{Q}$ order also display uniform (i.e., $q=0$) magnetism. In the case of an iCDW involving the same vHs, the latter corresponds to ferromagnetic (FM) order with spins on the Sb sites pointing out of the plane. Conversely, in the case of iCDW involving different vHs, the \QQ{} mixed phase displays no net dipolar moment. Instead, we find that it gives rise to exotic types of uniform magnetism, such as octupolar, toroidal, and even monopolar magnetic order. We show that each of these subsidiary orders couples uniquely to a combination of external magnetic, electric, and strain fields, displaying characteristic magneto-striction and multiferroic properties that can be detected experimentally. 

As for the $2\mathbf{Q}$-$1\mathbf{Q}$ mixed iCDW-rCDW configurations, although they do not display any type of uniform magnetic order, they spontaneously break the threefold rotational symmetry of the lattice and give rise to an orthorhombic distortion. This happens regardless of whether the iCDW is made out of electronic states from the same or different vHs. Finally, we extend our analysis to the case of an iCDW modulated along the $c$-axis and discuss the broad implications of our results for the identification of the complex charge-order patterns realized in the \AVS{} compounds, as well as their interplay with superconductivity.

This paper is organized as follows: In Sec.~\ref{sec:bloch_states}, we analyze the orbital and sublattice characters of the Bloch states at the van Hove singularities at the M point and classify these according to the irreps of the $P6/mmm$ space group. The candidate charge orders that can arise from the occupied vHs Bloch states are introduced in Sec.~\ref{sec:occupied}. These come in both intra- (Sec.~\ref{subsec:intra}) and inter-orbital (Sec.~\ref{subsec:inter}) varieties, each giving rise to distinct types of charge order with different properties. In Sec.~\ref{sec:free_energy}, we introduce the coupled Landau free energy of real and imaginary charge orders and argue that a mixed configuration -- either a \QQ{} or a \Q{} phase -- is generally favored. Section~\ref{sec:implications} presents the experimental signatures of the mixed charge ordered phases. In Sec.~\ref{sec:unoccupied}, we discuss the impact of the unoccupied vHs Bloch states. Finally, Sec.~\ref{sec:conclusions} contains our conclusions. Additional details of the first-principles calculations are show in Appendix \ref{app:DFT}, whereas details of the classification of the Bloch states in terms of space group irreps are included in Appendix~\ref{app:space_group_irreps}.

\section{Orbital and sublattice character of the van Hove singularities}\label{sec:bloch_states}

We start by employing group theory to elucidate the symmetry properties of the low-energy electronic states near the vHs. Interactions involving these states have been proposed to give rise to different types of CDW and SC order \cite{Park2021,Lin2021,Denner2021}. Note that a similar type of analysis was previously done in Ref.~\onlinecite{Park2021}. In this paper, besides recovering some of the results of Ref.~\onlinecite{Park2021}, we consider additional types of iCDW order and mixed iCDW-rCDW configurations.
\begin{figure}[t]
    \includegraphics[width=\columnwidth]{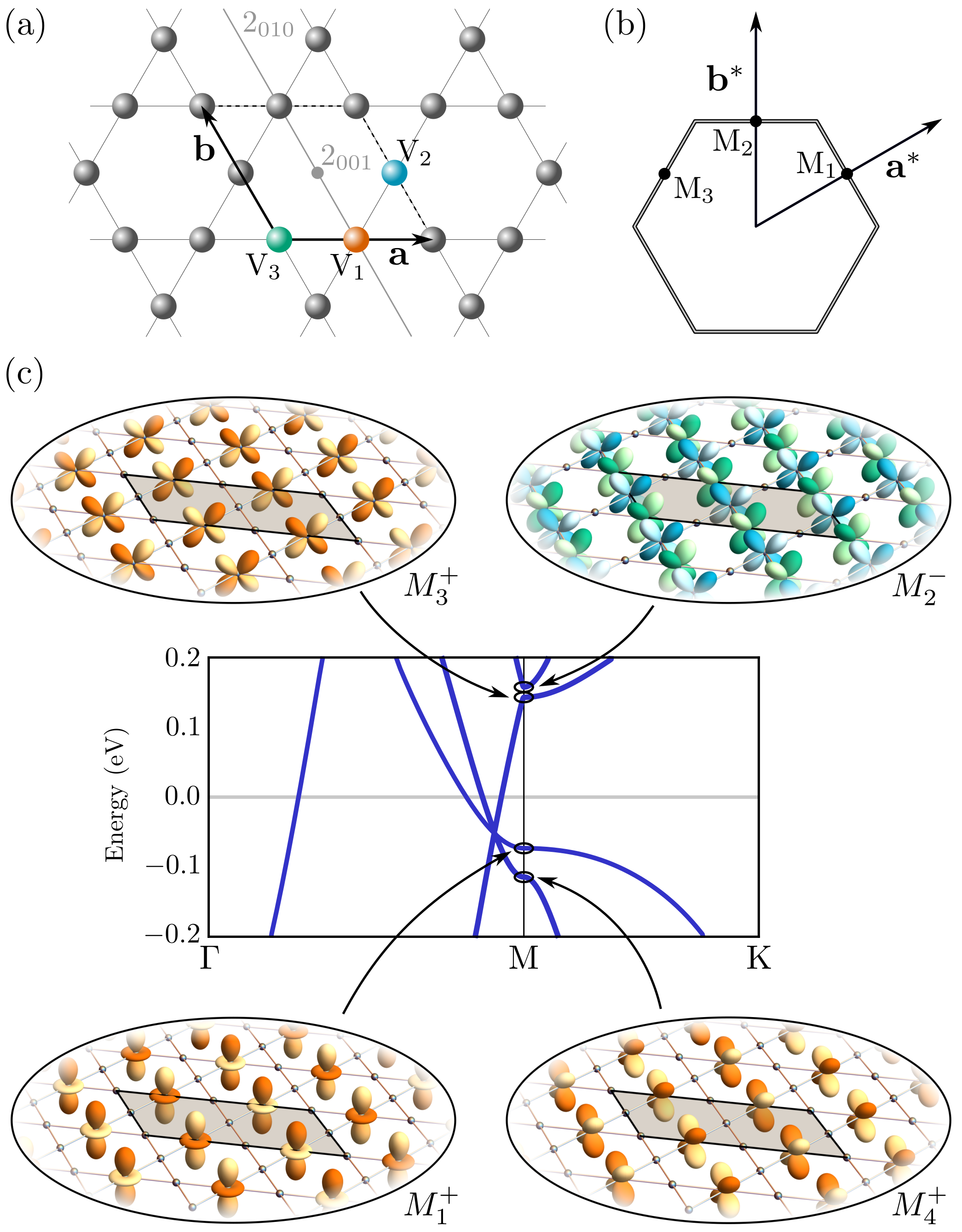}\\
    \caption{\label{fig:orbitals_and_bands} (a) Vanadium kagome layer of \AVS{} including the choice of lattice vectors, $\mbf{a}$ and $\mbf{b}$, and the three distinct V sublattice sites, V$_1$, V$_2$, and V$_3$ (in orange, blue, and green respectively). The generators of $D_{2h}$, which is the V site-symmetry point group, are denoted in gray and consist of two axes of two-fold rotations, $2_{001}$ and $2_{010}$, as well as inversion. (b) Hexagonal BZ of the unit cell shown in (a) with reciprocal lattice vectors $\mbf{a}^{\ast}$ and $\mbf{b}^{\ast}$ and the three symmetry-related M points denoted by M$_1$, M$_2$ and M$_3$.
    (c) Band structure along the $\Gamma-\rm{M}-\rm{K}$ direction obtained from DFT calculations for CsV$_3$Sb$_5$ and the associated Bloch states near M$_1$. Four vHs (two occupied and two unoccupied) arise from the saddle points at the M point, whose Bloch states are dominated by the $d_{z^2}$, $d_{xz}$, and $d_{yz}$ V orbitals (insets). Note that the sign of each orbital is modulated along M$_1$. The extended $2\times 1$ unit cell is denoted by the gray rectangle in the insets. At three of the M$_1$ vHs, the Bloch states are composed of orbitals located at $V_1$ ($p$-type vHs). The Bloch state at the fourth vHs, which is farthest from the Fermi level, is composed of orbitals located at the $V_2$ and $V_3$ sites ($m$-type vHs). The irreps of each Bloch state are indicated in the insets.}
\end{figure}

At room temperature, the \AVS{} compounds belong to the crystallographic space group $P6/mmm$, which has a simple hexagonal BZ. In Fig.~\ref{fig:orbitals_and_bands}(a) we illustrate the vanadium atoms of the kagome layer, with the three distinct vanadium sublattice sites highlighted by different colors. The vHs correspond to the saddle points of the band dispersion located at the M points of the BZ. There are three symmetry-related M points, which we denote by M$_i$ and describe by the momenta $\mbf{Q}_i=(\pm \pi,\tfrac{\pi}{\sqrt{3}},0),(0,\tfrac{2\pi}{\sqrt{3}},0)$, with $i=1,2,3$ [see Fig. \ref{fig:orbitals_and_bands}(b)]. As shown in Fig. \ref{fig:orbitals_and_bands}(c), electronic structure calculations using density-functional theory (DFT) predict two vHs below the Fermi level and two vHs above the Fermi level at the M point in the \AVS{} materials~\cite{Ortiz2019New,Tan2021,Jiang2021Unconventional,Wu2021Nature,Kang2022Twofold,Gu2022Gapless,Jeong2021Crucial}. This is consistent with results from angle-resolved photo-emission spectroscopy (ARPES)~\cite{Kang2022Twofold,Hu2022Rich}. We note that the band dispersions associated with the two vHs located above the Fermi level undergo an avoided crossing before the M point is reached, which masks the saddle points. While the electronic structure shown in Fig.~\ref{fig:orbitals_and_bands}(c) is obtained for CsV$_3$Sb$_5$, the general features are valid for all the \AVS{} compounds. In particular, the analysis of the wave-functions presented below does not depend on the specific compound in question.

Let us first analyze the wave-functions of the vHs below the Fermi level, i.e. the occupied vHs. The key result is that these saddle points have atomic contributions from only one of the three vanadium atoms in the unit cell. In other words, labeling the three distinct V atoms in real space by V$_i$ [Fig.~\ref{fig:orbitals_and_bands}(a)], the saddle points at a particular M$_i$ point in momentum space below the Fermi level [Fig.~\ref{fig:orbitals_and_bands}(b)] arise from the dispersion of orbitals at V$_i$ only. Thus, in the notation of, e.g., Refs.~\onlinecite{Kiesel2012,Kang2022Twofold}, both saddle points correspond to a $p$-type (i.e., ``pure") vHs. To highlight the fact that these saddle points are located below the Fermi level, here we denote them as $p_-$-type vHs.

The main difference between these two vHs is that they have different vanadium orbital characters. To show that, we classify them in terms of the irreducible representations (irrep) of the space group $P6/mmm$ at the M point, which are labelled by $M_{\alpha}^{\pm}$ with $\alpha = 1,\ldots,4$ (see also Table~\ref{tab:irrep}). Note that italic $M$ refers to the irrep whereas regular M refers to the BZ point. While the state at the occupied saddle point at M closest to (but below) the Fermi level transforms as the $M_1^+$ irrep, the state at the lower one transforms as the $M_4^+$ irrep. To connect these irreps to the V $d$-orbitals, we note that the latter can be classified according to the irreps of the site-symmetry point group of the V site, which is $D_{2h}$. Considering first the fully symmetric $A_{g}$ orbitals, corresponding to the $d_{z^2}$ and $d_{x^2-y^2}$ orbitals, we can use group theory to establish the irreps of the bands induced by them~\cite{Elcoro2017}:
\begin{equation}
    A_g \uparrow P6/mmm \sim M_1^+ \oplus M_3^- \oplus M_4^-\,.\label{eq:induced_rep}
\end{equation}
In this notation, the $z$-axis is parallel to the crystallographic $c$-direction, whereas the $x$-axis can be chosen parallel or perpendicular to the $b$-direction for $V_1$. Note that even though all three vanadium atoms are symmetry-equivalent, they each have a different local coordinate system that differ by a rotation around the $z$-axis. Hereafter, we use local coordinate axes such that the $x$-axes point towards the center of the hexagon formed by vanadium atoms. Therefore, from Eq. (\ref{eq:induced_rep}), we conclude that the occupied saddle-point nearest to the Fermi level at the M point has $A_g$ orbital character. Our DFT calculations (details in Appendix \ref{app:DFT}) reveal that the $d_{z^2}$ orbital provides the leading contribution, whereas $d_{x^2-y^2}$ gives the sub-leading one. The fact that the orbitals of only one of the three V atoms contributes to this saddle point is illustrated in Fig.~\ref{fig:orbitals_and_bands}(c), where we show for simplicity only the dominant $d_{z^2}$ orbital. Note that the appearance of only $d_{z^2}$ and $d_{x^2-y^2}$ as orbital states is a consequence of using a local coordinate system for each V atom. As the local coordinate system is different at each V atom, the lobes of the $d_{x^2-y^2}$ states are also oriented differently at each site. Hence, the $d_{x^2-y^2}$ states cannot be defined with respect to a global coordinate system. In this case, the states at two of the V atoms would become superpositions of $d_{x^2-y^2}$ and $d_{xy}$, defined with respect to the global coordinate system, in addition to the dominant $d_{z^2}$ orbital.

The second occupied saddle point at M closest to the Fermi level arises from a $B_{2g}$ orbital (or $B_{3g}$ depending on the choice of the $x$-axis), which corresponds to the $d_{xz}$ orbital in the aforementioned local coordinate system. If a single global coordinate system is used for all three vanadium atoms in the unit cell, then the $B_{2g}$ orbitals would correspond to a linear superposition of $d_{xz}$ and $d_{yz}$ orbitals
(see Appendix~\ref{app:space_group_irreps} for details). In a manner similar to Eq.~\eqref{eq:induced_rep}, the $B_{2g}$ orbital also induces three bands near M, one of which transforms precisely as $M_4^+$. This band also has a particularly simple wave-function with contributions from only a single V atom in the unit cell, as shown in Fig.~\ref{fig:orbitals_and_bands}(c).

\begin{figure}
    \includegraphics[width=0.95\columnwidth]{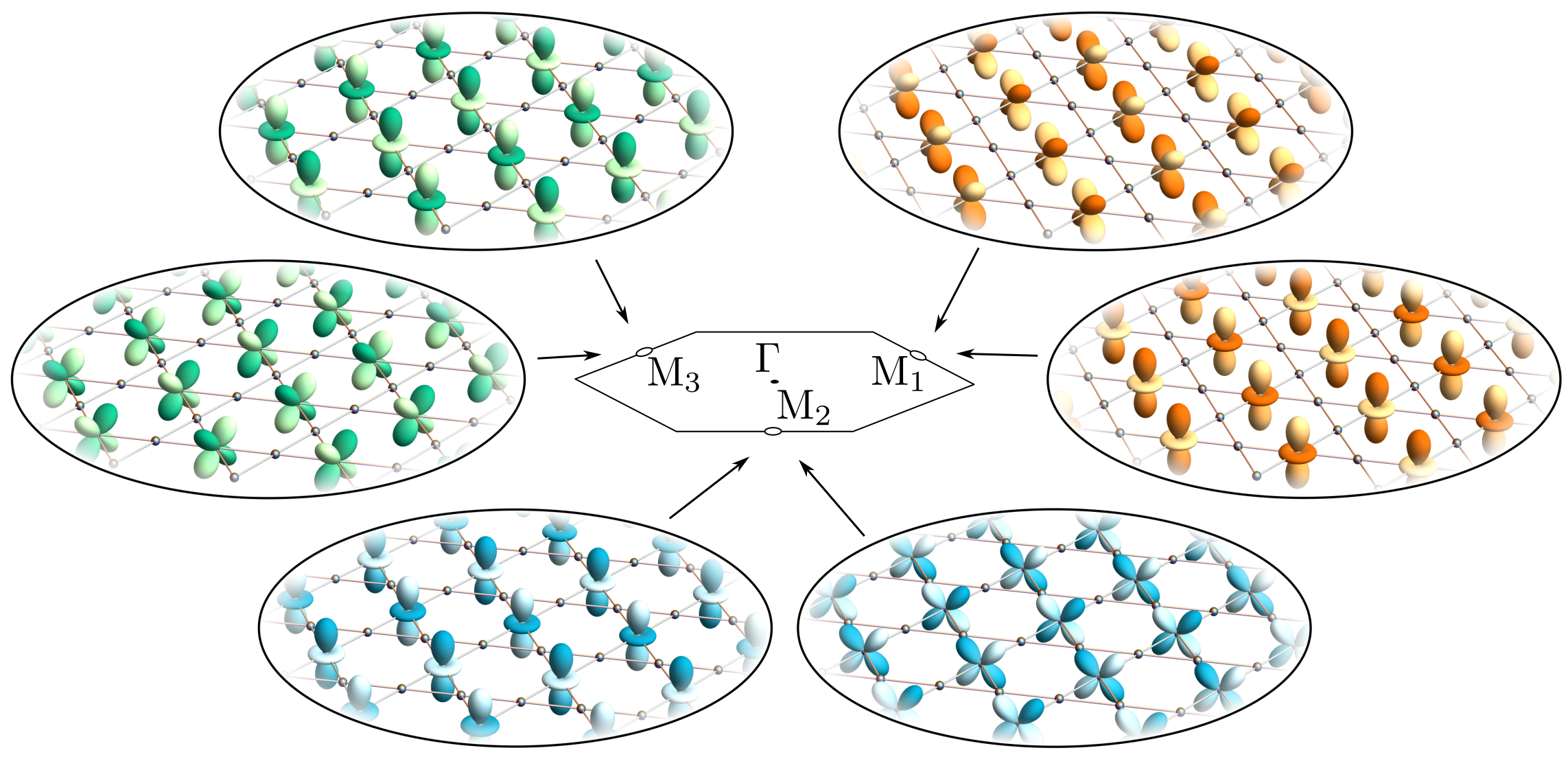}
    \caption{\label{fig:states_near_M} Orbital and V-sublattice character of the Bloch states describing the two \emph{occupied} vHs closest to (but below) the Fermi level at each of the three M points. The phase of the orbital is modulated along a different direction depending on the BZ point, M$_1$, M$_2$, and M$_3$. The orbital weight for a given M point is concentrated on a distinct V site resulting in $p$-type vHs. Moreover, the V $d_{xz}$-orbital is defined with respect to a local coordinate system, as explained in the main text.}
\end{figure}

Figure~\ref{fig:states_near_M} summarizes the main results of the group theory analysis performed in this section for the Bloch states of the occupied saddle points at the three different M points that are closest to (but below) the Fermi level. At each M point, the wave-functions of a given saddle point are located on different V atoms. Moreover, the spectral weight of each of the two vHs at a given M point is dominated by a different type of V $d$-orbital, denoted here by $A_g$ ($d_{z^2}$, $d_{x^2-y^2}$) and $B_{2g}$ ($d_{xz}$). As we discuss in Sec. \ref{sec:occupied}, the different symmetry properties of the two occupied van Hove singularities have important implications for the types of charge-order that can arise in the \AVS{} kagome metals. 

We now move to the two unoccupied vHs whose energies are above the Fermi energy. While they seem more distant from the Fermi level as the pair of occupied vHs, the relative energy difference is small enough that it is prudent to also consider them in the analysis. The unoccupied vHs closest to the Fermi level has the same $p$-type structure as the occupied vHs -- in this case, we denote it $p_+$-type to emphasize that it is located above the Fermi level. As illustrated in Fig.~\ref{fig:orbitals_and_bands}(c), its wave-function is dominated by the (local) $d_{yz}$ orbitals from a single vanadium site. In contrast to the occupied vHs, the lobes of these $B_{3g}$ orbitals point towards the V-V bonds. As a result, this Bloch state transforms as a different irrep than the two occupied ones, namely $M_3^+$.

\begin{figure}
    \includegraphics[width=0.95\columnwidth]{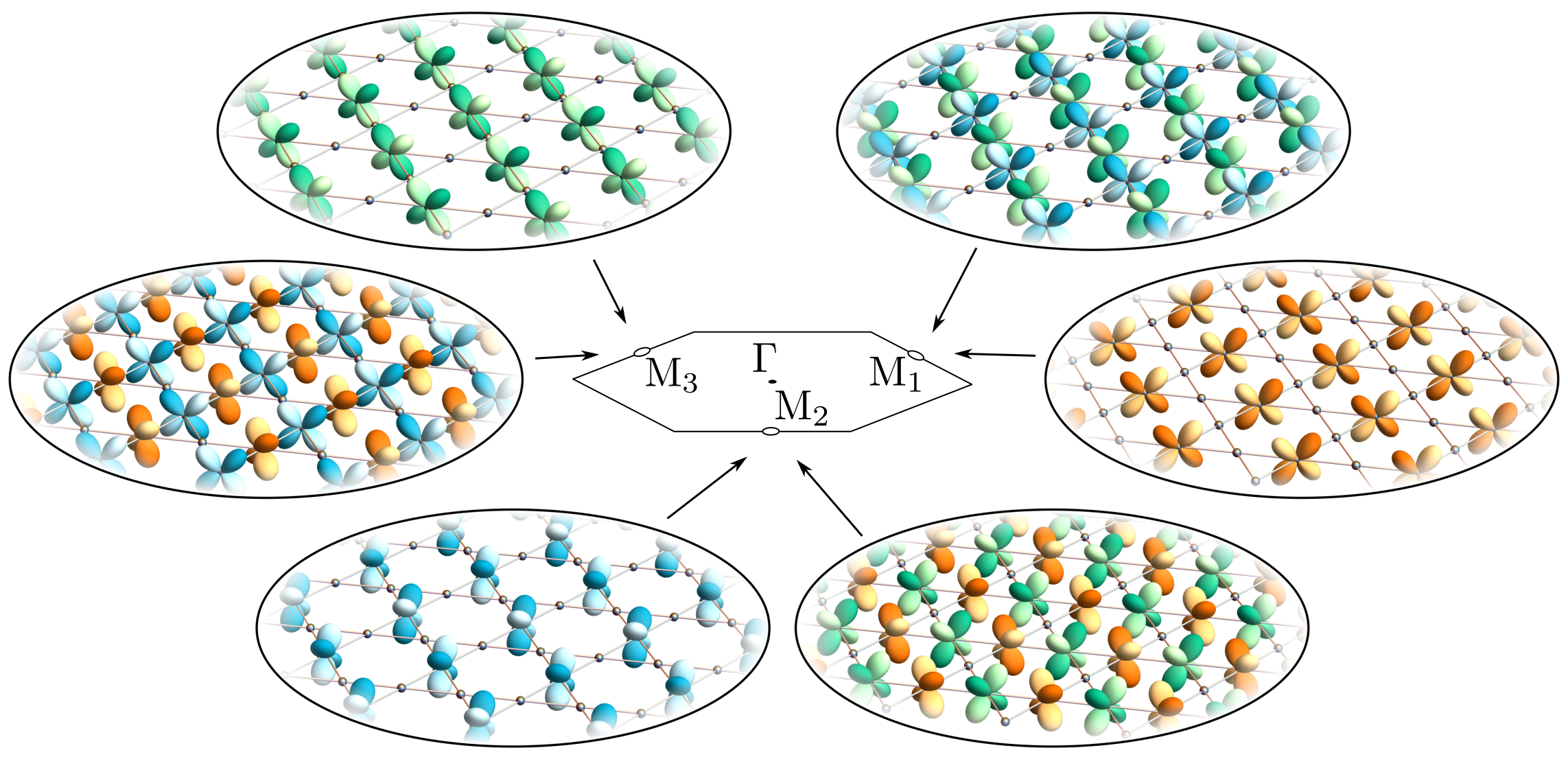}
    \caption{\label{fig:states_near_M_unoccupied} Orbital and V-sublattice character of the Bloch states describing the two \emph{unoccupied} vHs above the Fermi level at each of the three M points. The phases of the orbitals are modulated along a different direction depending on the BZ point, M$_1$, M$_2$, and M$_3$. The $d_{yz}$ orbitals contributing to the Bloch state associated with the unoccupied vHs closer to the Fermi level arise from a single V site. This is similar to the situation in Fig.~\ref{fig:states_near_M}, thus resulting in a $p$-type vHs. On the other hand, the $d_{xz}$ orbitals contributing to the other unoccupied vHs arise from the other two V sites, leading to an $m$-type vHs.}
\end{figure}

The wave-function of the unoccupied vHs that is farthest from the Fermi energy in Fig.~\ref{fig:orbitals_and_bands} has a completely different sublattice structure from the other vHs discussed so far. As shown in Fig.~\ref{fig:orbitals_and_bands}(c), this wave-function is also dominated by $d_{xz}$ orbitals in the local coordinate system, but from the other two vanadium sites. More specifically, at the M$_i$ point, the vHs wave-function has sublattice contributions from the vanadium sites V$_j$ and V$_l$, where $(i,j,l)$ is a permutation of $(1,2,3)$. Using the nomenclature of Refs.~\onlinecite{Kiesel2012,Kang2022Twofold}, this is an $m$-type (i.e., ``mixed") vHs; since there is no $m$-type vHs below the Fermi energy in our low-energy model, we do not include the subscript ${+}$. More importantly, the Bloch state of this $m$-type vHs transforms as the $M_2^-$ irrep of the space group. The minus sign in the superscript is a consequence of the fact that this combination of orbitals is odd under an inversion operation with respect to the center of the hexagon formed by V sites. As we will discuss later in Sec. \ref{sec:unoccupied}, the distinct symmetry properties of this particular vHs have crucial implications for the types of iCDW that it can generate. In Fig. \ref{fig:states_near_M_unoccupied}, we summarize the orbital and sublattice character of the Bloch states corresponding to the pair of unoccupied vHs.

\section{Candidate charge orders: occupied van Hove singularities} \label{sec:occupied}

Because they are the ones closest to the Fermi level, we start our analysis by considering first only the two occupied $p_{-}$-type vHs at the M point. These are made out of different vanadium orbitals from the same V sublattice. The role of the unoccupied vHs will be considered in Sec.~\ref{sec:unoccupied}. In our approach, we depart from the electronic Bloch states obtained in Sec.~\ref{sec:bloch_states}, classify the possible intra- and inter-orbital charge order parameters, and discuss their real-space realizations. We note that a similar analysis was performed in Ref.~\onlinecite{Park2021} considering only one of the saddle points. 

Our focus is on the rCDW and iCDW order parameters with wave-vector $\mbf{Q}_{M_i}$, which we denote by $N_i$ and $\Phi_i$, respectively, following the notation of Ref.~\onlinecite{Park2021}. Importantly, these order parameters must transform as one of the $M_{\alpha}^{\pm}$ irreps of the $P6/mmm$ space group. Because the irreps associated with rCDW and iCDW transform in different ways under the time reversal operation (even and odd, respectively), they are distinct order parameters with different symmetry properties. Formally, the irreps $M_{\alpha}^{\pm}$ are all real and three-dimensional, arising from the three vectors in the star of M. Hence, order parameters transforming as these irreps have three components. As a consequence, the rCDW and iCDW order parameters are independent of each other, and do not transform as the real and imaginary parts of a single complex CDW order parameter as this would have six components. 

\subsection{Intra-orbital rCDW and iCDW}\label{subsec:intra}

Let $c^\dagger_\mbf{k\sigma}$ and $d^{\dagger}_\mbf{k\sigma}$ denote the creation operator of an energy eigenstate near the occupied $M_1^+$ and $M_4^+$ vHs, respectively. Here, $\mbf{k}$ denotes momentum and $\sigma$, spin. Because the $M_1^+$ and $M_4^+$ saddle points are composed of different types of orbitals ($A_g$ and $B_{2g}$, respectively), the allowed intra-orbital charge order parameters with wave-vector $\mbf{Q}_{i}$ are those that combine fermions of the same species, i.e. fermions from the same type of saddle-point at two different M points. 

As discussed above, there are two independent types of charge order: rCDW and iCDW. The intra-orbital rCDW order parameters are given by (see also Ref.~\onlinecite{Park2021}):
\begin{align}
    N^c_{i} &= \sum_{\mbf{k}\sigma} \langle c^{\dagger}_{\mbf{k}+\mbf{Q}_{j}\sigma} c^{\phantom{\dagger}}_{\mbf{k}+\mbf{Q}_{l}\sigma} +\text{H.c.} \rangle\,, \label{eq:Mc_def} \\
    N^d_{i} &= \sum_{\mbf{k}\sigma} \langle d^{\dagger}_{\mbf{k}+\mbf{Q}_{j}\sigma} d_{\mbf{k}+\mbf{Q}_{l}\sigma} +\text{H.c.} \rangle\,. \label{eq:Md_def}
\end{align}
In the equations above and in all definitions of the CDW order parameters in this paper, we use the convention that $(i,j,l)$ is a permutation of $(1,2,3)$. The subscript $i=1,2,3$ denotes the three components of the order parameter, associated with the three wave-vectors $\mbf{Q}_{M_i}$ in the star of M. Similarly, the iCDW order parameters are
\begin{align}
    \Phi^{c}_{i} &= i \sum_{\mbf{k}\sigma} \langle c^{\dagger}_{\mbf{k}+\mbf{Q}_{j}\sigma} c_{\mbf{k}+\mbf{Q}_{l}\sigma} - \text{H.c.} \rangle\,, \label{eq:Phic_def} \\
    \Phi^{d}_{i} &= i \sum_{\mbf{k}\sigma} \langle d^{\dagger}_{\mbf{k}+\mbf{Q}_{j}\sigma} d_{\mbf{k}+\mbf{Q}_{l}\sigma} - \text{H.c.} \rangle\,. \label{eq:Phid_def}
\end{align}
Note that, despite the ``imaginary" denomination, the iCDW order parameter is real-valued. As we emphasize below, the distinction between rCDW and iCDW stems from the fact that the first corresponds to charge disproportionation at sites and/or bonds and the second, to loop-currents.

The symmetry properties of the order parameters $N_i^{\mu}$ and $\Phi_i^{\mu}$, with $\mu=c,d$, can be directly obtained from the symmetry properties of the creation and annihilation operators. The latter, in turn, are determined by the Bloch states derived in Sec.~\ref{sec:bloch_states}, which transform as either $M_1^+$ (for $\mu=c$ fermions) or $M_4^+$ (for $\mu=d$ fermions). Details of the derivation can be found in Appendix~\ref{app:space_group_irreps}. For the rCDW case, we find that both intra-orbital order parameters $N_i^{\mu}$ transform as the $M_1^+$ irrep of the $P6/mmm$ group. Since the point symmetry part of the little group at M is $D_{2h}$, this means that each of the three components of $N_i^{\mu}$ transforms trivially under the operations of $D_{2h}$ (with rotation axes aligned differently for each wave-vector), as shown in Table \ref{tab:irrep} -- although they break the translational symmetry along the $i$-th direction. Such an rCDW order corresponds to V-V bond distortions, and is consistent with experimental evidence for the presence of an $M_1^+$ lattice distortion concomitant with the onset of charge order~\cite{Ratcliff2021,Uykur2022Optical,Wu2022Charge}. The triple-$\mathbf{Q}$ rCDW order, which corresponds to the condensation of all three components $N_i^{\mu}$ with equal amplitudes, gives rise to either the star-of-David or tri-hexagonal bond-order configurations depending on the sign of $N_1 N_2 N_3$ \cite{Christensen2021}.

\begin{table}
\begin{tabular}{@{\hskip .2in}c@{\hskip .2in}|@{\hskip .2in}c@{\hskip .2in}@{\hskip .2in}c@{\hskip .2in}@{\hskip .2in}c@{\hskip .2in}}
        & $2_{001}$     & $2_{010}$     & $\bar{1}$\\
\hline
$M_1^{\pm}$ & $+1$             & $+1$             & $\pm 1$     \\
$M_2^{\pm}$ & $+1$             & $-1$            & $\pm 1$     \\
$M_3^{\pm}$ & $-1$            & $+1$             & $\pm 1$     \\
$M_4^{\pm}$ & $-1$            & $-1$            & $\pm 1$     \\
\end{tabular}
        \caption{\label{tab:irrep} Characters of the irreps of the little group of the M$_1$ point. We only list the characters for the three generators of the little group, and these irreps correspond to the space group irreps by the same name. Note that even though the little group irreps are one-dimensional, the corresponding space group irreps are three-dimensional, because the star of $M$ has three distinct wave-vectors.}
\end{table}

\begin{figure}
    \includegraphics[width=\columnwidth]{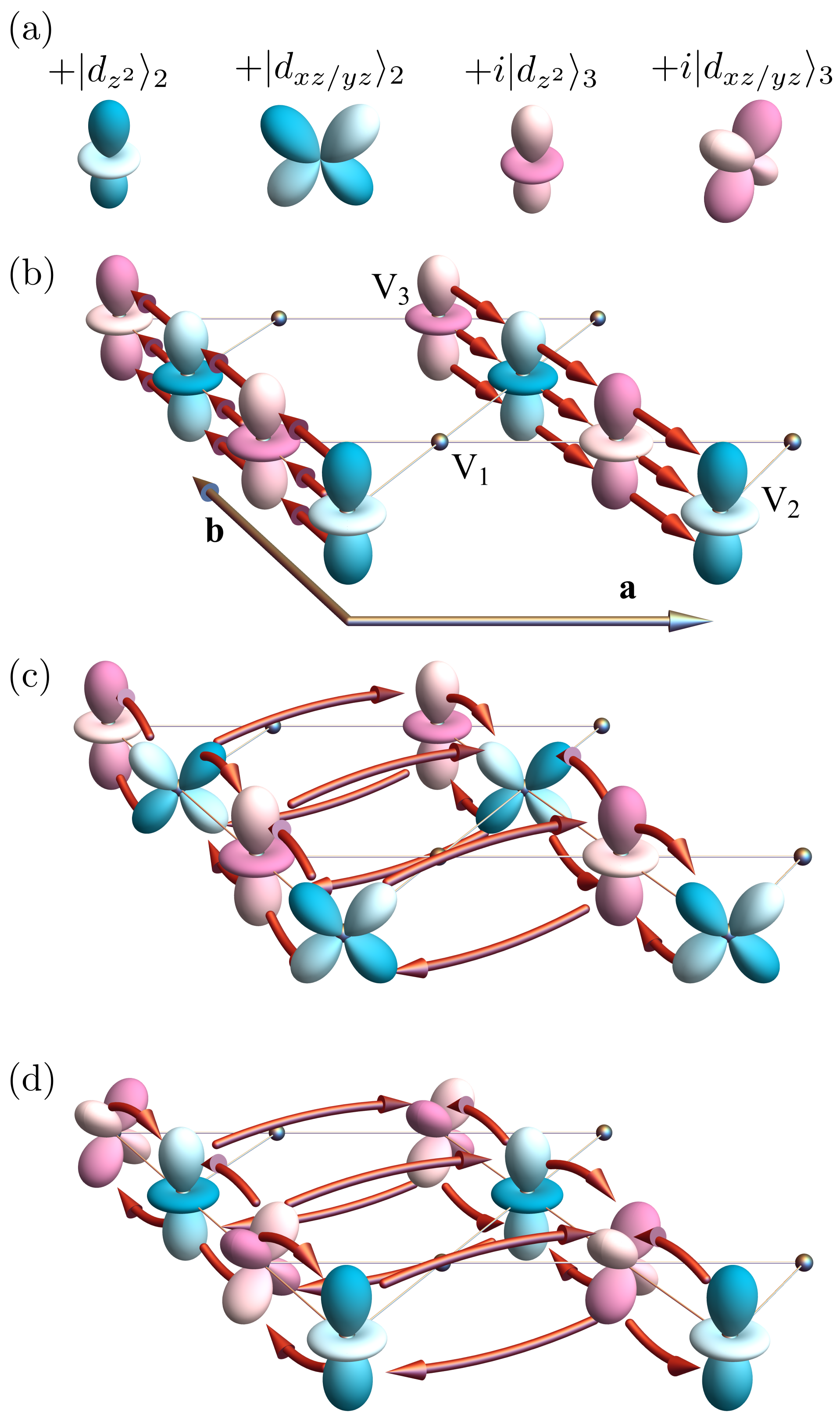}
    \caption{\label{fig:orbital_hoppings} Schematics of the different constituents of the possible iCDW phases involving the occupied vHs, in terms of currents involving the $A_g$ orbitals ($d_{z^2}$) and the $B_{2g}$ orbitals ($d_{xz}$). (a) defines the phase of the orbitals, and (b)-(d) refer to iCDW order along M$_1$, where the Bloch states are centered on the $V_2$ and $V_3$ sites. While (b) corresponds to the $mM_2^+$ intra-orbital iCDW phase, the $mM_3^+$ and $mM_4^+$ inter-orbital iCDW phases are the symmetric and antisymmetric combinations of (c) and (d), respectively.}
\end{figure}

As for the iCDW case, we find that the order parameters $\Phi_i^\mu$ transform as the $mM_2^+$ irrep of $P6/mmm$. Here, $m$ denotes the fact that the irrep is odd under time-reversal symmetry. Additionally, it breaks the lattice translational symmetry and the in-plane two-fold lattice rotation axis, as indicated in Table~\ref{tab:irrep}. In Fig.~\ref{fig:orbital_hoppings}(b), we illustrate $\Phi^{c}_1$ as hopping between $d_{z^2}$ orbitals centered on the $V_2$ and $V_3$ sites. The different colors denote a relative phase of $\pi/2$ between orbitals on neighboring V atoms; it is such a phase difference that defines the current direction. In this particular case, the translational symmetry is broken along the crystallographic $a$-axis, as expected for the wave-vector $\mbf{Q}_{1}$. A similar current pattern involving the $B_{2g}$ orbitals also exists simultaneously to the pattern shown in Fig.~\ref{fig:orbital_hoppings}(b). Indeed, since $\Phi^{c}_1$ and $\Phi^{d}_1$ transform as the same irrep, one order necessarily triggers the other one.

\begin{figure*}
    \includegraphics[width=\textwidth]{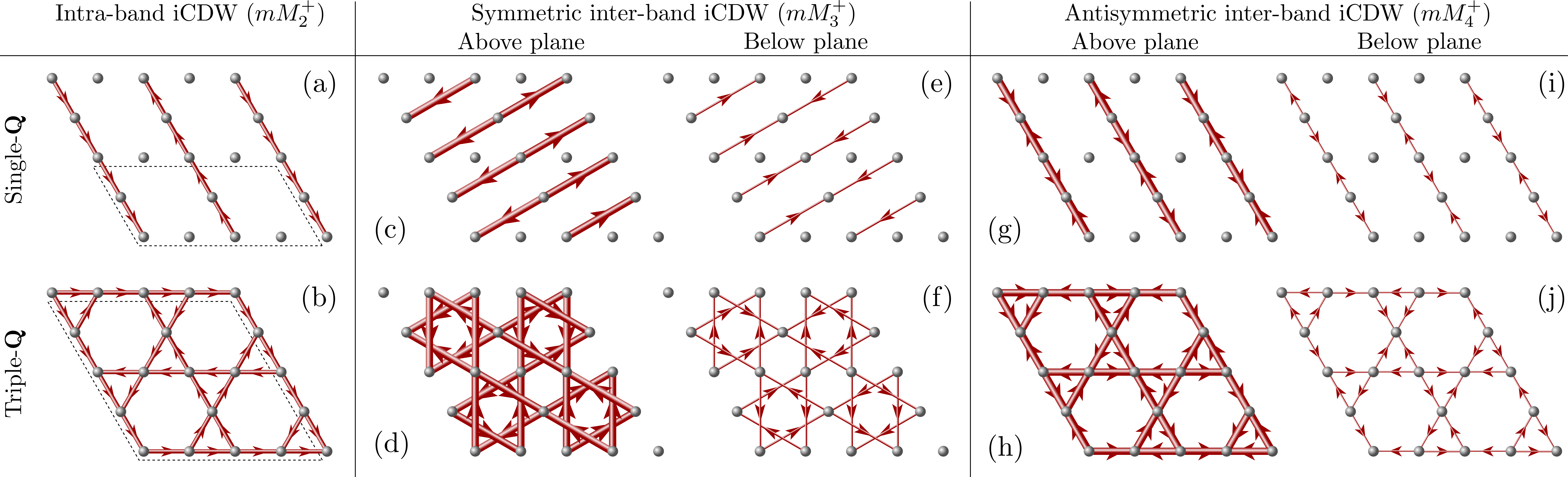}
    \caption{\label{fig:iCDW_phases}Illustration of the single-$\mbf{Q}$ (upper panels) and triple-$\mbf{Q}$ (lower panels) configurations associated with each of the three distinct types of iCDW order arising from the occupied vHs. The latter are labeled here by the M-point irrep of the $P6/mmm$ space group according to which they transform. The intra-orbital iCDW order, transforming as $mM_2^+$ [(a)-(b)], gives rise to currents in the kagome plane. The two types of inter-orbital iCDW order, $mM_3^+$ [(c)-(f)] and $mM_4^+$ [(g)-(j)], promote counter-propagating currents above and below the kagome planes, as discussed in the main text. Note that for the $mM_3^+$ and $mM_4^+$ cases, the configurations above and below the plane coexist. The dashed lines in (a) and (b) denote the $2\times1$ and $2\times2$ unit cells of the single-$\mbf{Q}$ and triple-$\mbf{Q}$ configurations. The thicker (thinner) arrows in (c)--(j) refer to currents above (below) the plane. Note that, due to the coexistence of the above- and below-plane configurations in (c)--(j), all states depicted here are charge-conserving.}
\end{figure*}

For a single-$\mathbf{Q}$ iCDW order involving only one V orbital, which is shown more schematically in Fig.~\ref{fig:iCDW_phases}(a), the loop currents close only at the edges of the system. In contrast, the triple-$\mbf{Q}$ iCDW configuration, consisting of an equal superposition of the three single-$\mbf{Q}$ order parameters $\Phi_i^\mu$ features the closed loops shown in Fig.~\ref{fig:iCDW_phases}(b). The properties of this type of order were previously considered in Refs.~\onlinecite{Park2021,Lin2021}.

\subsection{Inter-orbital rCDW and iCDW}\label{subsec:inter}

Inter-orbital charge orders are constructed by combining fermions from different types of occupied saddle points ($M_1^+$ and $M_4^+$) at two different M points. This case is more involved than the intra-orbital one, as only symmetric ($s$) and antisymmetric ($a$) combinations of the inter-band bilinears transform as irreps of the little group (see Appendix \ref{app:space_group_irreps}). For the rCDW case, we find
\begin{align}
        N_{i}^s &= \sum_{\mbf{k}\sigma} \langle
        c_{\mbf{k}+\mbf{Q}_{j}\sigma}^\dagger d_{\mbf{k}+\mbf{Q}_{l}\sigma} + c_{\mbf{k}+\mbf{Q}_{l}\sigma}^\dagger d_{\mbf{k}+\mbf{Q}_{j}\sigma} +\text{H.c.}
        \rangle\,, \label{eq:Ms_def} \\ 
        N_{i}^a &= \sum_{\mbf{k}\sigma} \langle
        c_{\mbf{k}+\mbf{Q}_{j}\sigma}^\dagger d_{\mbf{k}+\mbf{Q}_{l}\sigma} - c_{\mbf{k}+\mbf{Q}_{l}\sigma}^\dagger d_{\mbf{k}+\mbf{Q}_{j}\sigma} +\text{H.c.}
        \rangle\,, \label{eq:Ma_def}
\end{align}
which transform as the $M_3^+$ and $M_4^+$ irreps, respectively. As mentioned above, experimental evidence, as well as DFT calculations, point to the rCDW order transforming as $M_1^+$ as the one realized experimentally~\cite{Ratcliff2021,Wu2022Charge,Tan2021,PRM_Subedi2022,Li2021Observation,Uykur2022Optical}. Therefore, we will not discuss the $M_3^+$ and $M_4^+$ rCDW orders further. In contrast, the properties of the possible iCDW order are not known from either experiments or first-principles calculations. Similar to the rCDW case, it is necessary to form symmetric and antisymmetric combinations of the fermion bilinears to construct proper iCDW order parameters:
\begin{align}
        \Phi_{i}^s &= i\sum_{\mbf{k}\sigma} \langle
        c_{\mbf{k}+\mbf{Q}_{j}\sigma}^\dagger d_{\mbf{k}+\mbf{Q}_{l}\sigma} + c_{\mbf{k}+\mbf{Q}_{l}\sigma}^\dagger d_{\mbf{k}+\mbf{Q}_{j}\sigma} - \text{H.c.}
        \rangle\,, \label{eq:iCDW_symmetric} \\
        \Phi_{i}^a &= i\sum_{\mbf{k}\sigma} \langle
        c_{\mbf{k}+\mbf{Q}_{j}\sigma}^\dagger d_{\mbf{k}+\mbf{Q}_{l}\sigma} - c_{\mbf{k}+\mbf{Q}_{l}\sigma}^\dagger d_{\mbf{k}+\mbf{Q}_{j}\sigma} - \text{H.c.}
        \rangle\,. \label{eq:iCDW_antisymmetric} 
\end{align}
We find that, while $\Phi_i^s$ transforms as the $mM_3^+$ irrep of $P6/mmm$, $\Phi_i^a$ transforms as $mM_4^+$. In terms of the symmetry operations of the little group $D_{2h}$, $\Phi_i^s$ breaks the two-fold rotational symmetry with respect to the out-of-plane axis, whereas $\Phi_i^a$ breaks also the two-fold rotational symmetry with respect to the in-plane axis [see Table \ref{tab:irrep} and Fig.~\ref{fig:orbitals_and_bands}(a)]. 

The loop current patterns generated by these types of iCDW order can be obtained by combining the two inter-orbital current patterns shown in Figs.~\ref{fig:orbital_hoppings}(c)-(d) according to Eqs. (\ref{eq:iCDW_symmetric}) and (\ref{eq:iCDW_antisymmetric}). The resulting single-$\mathbf{Q}$ symmetric combination ($mM_3^+$), corresponding to $\Phi_1^s$, is shown schematically in Figs.~\ref{fig:iCDW_phases}(c) and (e), and connects the $V_2$ and $V_3$ sites on opposite sides of the hexagon. Notably, the current density associated with this iCDW order vanishes on the kagome layer and thus does not lead to a current on the V atoms. However, net currents are induced both above and below the kagome layer, which must necessarily involve the apical Sb atoms. Note that the simultaneous presence of currents above and below the plane implies that charge is conserved for these configurations as well. The fact that there is no current on the V atoms despite the iCDW order parameter being constructed from operators of V atomic orbitals can be understood from the structure of the orbitals in question. Indeed, the $mM_3^+$ order parameter mixes orbitals with and without a node on the kagome plane itself, as shown in Figs.~\ref{fig:orbital_hoppings}(c) and (d). As a consequence, such an iCDW phase is only possible in a multi-orbital model. The triple-$\mbf{Q}$ configuration of the $mM_3^+$ iCDW is shown in Figs.~\ref{fig:iCDW_phases}(d) and (f), and consists of counter-circulating loops above and below the plane, which coincide in a single hexagon. In contrast to the triple-$\mbf{Q}$ $mM_2^+$ iCDW configuration, it does not give rise to a dipole moment. We will see this manifested in Sec.~\ref{sec:implications} as well. 

\begin{table}
    \centering
    \begin{tabular}{C{0.06\columnwidth}C{0.82\columnwidth}C{0.06\columnwidth}}
         Order & Order parameter(s) & Irrep \\
         \hline
         \hline
         rCDW & $\displaystyle N_i^c=\sum_{\mbf{k}\sigma} \langle c^{\dagger}_{\mbf{k}+\mbf{Q}_{j}\sigma} c_{\mbf{k}+\mbf{Q}_{l}\sigma} +\text{H.c.} \rangle $ \newline 
          $\displaystyle N_i^d=\sum_{\mbf{k}\sigma} \langle d^{\dagger}_{\mbf{k}+\mbf{Q}_{j}\sigma} d_{\mbf{k}+\mbf{Q}_{l}\sigma} +\text{H.c.} \rangle$ & $M_1^+$ \\
          & $\displaystyle N_{i}^s = \sum_{\mbf{k}\sigma} \langle
        c_{\mbf{k}+\mbf{Q}_{j}\sigma}^\dagger d_{\mbf{k}+\mbf{Q}_{l}\sigma} + c_{\mbf{k}+\mbf{Q}_{l}\sigma}^\dagger d_{\mbf{k}+\mbf{Q}_{j}\sigma} +\text{H.c.}
        \rangle$ & $M_3^+$ \\
          & $\displaystyle N_{i}^a = \sum_{\mbf{k}\sigma} \langle
        c_{\mbf{k}+\mbf{Q}_{j}\sigma}^\dagger d_{\mbf{k}+\mbf{Q}_{l}\sigma} - c_{\mbf{k}+\mbf{Q}_{l}\sigma}^\dagger d_{\mbf{k}+\mbf{Q}_{j}\sigma} +\text{H.c.}
        \rangle$ & $M_4^+$ \\
        \hline
         iCDW & $\displaystyle \Phi_i^c=i\sum_{\mbf{k}\sigma} \langle c^{\dagger}_{\mbf{k}+\mbf{Q}_{j}\sigma} c_{\mbf{k}+\mbf{Q}_{l}\sigma} -\text{H.c.} \rangle$ \newline $\displaystyle \Phi_i^d=i\sum_{\mbf{k}\sigma} \langle d^{\dagger}_{\mbf{k}+\mbf{Q}_{j}\sigma} d_{\mbf{k}+\mbf{Q}_{l}\sigma} -\text{H.c.} \rangle$ & $mM_2^+$ \\
          & $\displaystyle \Phi_{i}^s = i\sum_{\mbf{k}\sigma} \langle
        c_{\mbf{k}+\mbf{Q}_{j}\sigma}^\dagger d_{\mbf{k}+\mbf{Q}_{l}\sigma} + c_{\mbf{k}+\mbf{Q}_{l}\sigma}^\dagger d_{\mbf{k}+\mbf{Q}_{j}\sigma} - \text{H.c.}
        \rangle$ & $mM_3^+$ \\
          & $\displaystyle \Phi_{i}^a = i\sum_{\mbf{k}\sigma} \langle
        c_{\mbf{k}+\mbf{Q}_{j}\sigma}^\dagger d_{\mbf{k}+\mbf{Q}_{l}\sigma} - c_{\mbf{k}+\mbf{Q}_{l}\sigma}^\dagger d_{\mbf{k}+\mbf{Q}_{j}\sigma} - \text{H.c.}
        \rangle$ & $mM_4^+$ \\
    \end{tabular}
    \caption{\label{tab:iCDW_order_parameters}Possible rCDW and iCDW order parameters constructed from the $c^\dagger_\mbf{k\sigma}$ and $d^{\dagger}_\mbf{k\sigma}$ fermionic operators. The latter refer to energy eigenstates state of the $M_1^+$ and $M_4^+$ occupied saddle points, respectively. $M_i^+$ are irreps of the little group of the M point of the space group $P6/mmm$. As explained in the text, the inter-orbital CDW orders are obtained as symmetric and antisymmetric combinations of the inter-orbital fermionic bilinears. In the equations above, $(i,j,l)$ is a permutation of $(1,2,3)$.}
\end{table}

Finally, the single-$\mbf{Q}$ and triple-$\mbf{Q}$ configurations associated with the antisymmetric ($mM_4^+$) iCDW order, described by $\Phi_1^a$, are depicted in Figs.~\ref{fig:iCDW_phases}(g)-(j). Like the $mM_3^+$ case, these configurations feature no net currents in the kagome layer itself. Moreover, the triple-$\mbf{Q}$ case also consists of counter-circulating loops above and below the plane that, once again, results in no uniform dipole moment.

In Table~\ref{tab:iCDW_order_parameters}, we summarize our results for the symmetry properties of the rCDW and iCDW order parameters formed out of fermions from the two vHs below the Fermi level.

\section{Mixed real and imaginary charge order configurations: occupied vHs} \label{sec:free_energy}

Experimentally, the condensation of a CDW order at $T_{\rm CDW}$ has been attributed to the softness of a specific $M_1^+$ phonon mode~\cite{Tan2021,Ratcliff2021,Li2021Observation,Uykur2022Optical,Wu2022Charge}. Since iCDW states transforming as $M_1^+$ (or $mM_1^+$) do not appear, this soft phonon has been interpreted as signalling the onset of an rCDW phase, corresponding to charge bond-order. However, this does not explain the observations of time-reversal symmetry breaking below $T_{\rm CDW}$ ~\cite{Kenney2021Absence,Mielke2022Time-reversal,Guguchia2022Tunable,Xu_MOKE}. Instead, these observations could be explained by an iCDW. To reconcile these two scenarios, it has been pointed out that a multi-$\mathbf{Q}$ iCDW necessarily triggers an rCDW due to the existence of trilinear terms in the free energy expansion of the coupled iCDW-rCDW order parameters \cite{Lin2021,Park2021}. As  a result, an iCDW instability may be enough to explain the observations of both a soft $M_1^+$ phonon mode and the spontaneous breaking of time-reversal symmetry below $T_{\rm CDW}$. 

Ref.~\onlinecite{Park2021} explored the case in which a triple-$\mathbf{Q}$ iCDW induces a triple-$\mathbf{Q}$ rCDW -- a mixed iCDW-rCDW configuration that we dub \QQ. As we show in this section, there is another viable mixed configuration in which a double-$\mathbf{Q}$ iCDW and a single-$\mathbf{Q}$ rCDW coexist -- which we denote \Q. To study the possible mixed iCDW-rCDW configurations, we construct the free energy of the high-symmetry phase by finding all possible polynomials of $N_i$ and $\Phi_i$ that remain invariant under the symmetry operations of the space group, as well as time reversal symmetry~\cite{Hatch2003INVARIANTS,StokesInvariants}. While we restrict $N_i$ to be a $M_1^+$ rCDW (i.e. an intra-orbital rCDW), we allow $\Phi_i$ to be any of three possible iCDW states constructed from the occupied vHs (see Table \ref{tab:iCDW_order_parameters}), i.e. intra-orbital ($mM_2^+$), symmetric inter-orbital ($mM_3^+$), and antisymmetric inter-orbital ($mM_4^+$). It turns out that, in all cases, the Landau free-energy expansion acquires the same form:
\begin{equation}
    \mathcal{F}_{\mathrm{CDW}} = \mathcal{F}_{\mathrm{r}} + \mathcal{F}_{\mathrm{i}} + \mathcal{F}_{\mathrm{i-r}}\,, \label{eq:Ftot}
\end{equation}
with the rCDW and iCDW free energies
\begin{align}
    \mathcal{F}_\mathrm{r} &= \frac{a_\mathrm{r}}{2}N^2 + \frac{\gamma_\mathrm{r}}{3}N_1 N_2 N_3 \nonumber \\ & +\frac{u_\mathrm{r}}{4}N^4 + \frac{\lambda_\mathrm{r}}{4} \left( N_1^2 N_2^2 + N_1^2 N_3^2 + N_2^2 N_3^2 \right) \,, \\
    \mathcal{F}_\mathrm{i} &= \frac{a_\mathrm{i}}{2}\Phi^2 + \frac{u_\mathrm{i}}{4}\Phi^4 + \frac{\lambda_\mathrm{i}}{4}\left(\Phi_1^2\Phi_2^2 + \Phi_1^2\Phi_3^2 + \Phi_2^2\Phi_3^2\right)\,, \label{eq:F_i}
\end{align}
and the coupling between them given by
\begin{align}
    \mathcal{F}_\mathrm{i-r} &= \frac{\gamma_\mathrm{ir}}{3}\left( N_1 \Phi_2 \Phi_3 + \Phi_1 N_2 \Phi_3 + \Phi_1 \Phi_2 N_3 \right) \nonumber \\ & + \frac{\kappa_\mathrm{ir}}{4} \left(N_1 N_2 \Phi_1 \Phi_2 + N_1 N_3 \Phi_1 \Phi_3 + N_2 N_3 \Phi_2 \Phi_3 \right) \nonumber \\
	& + \frac{\lambda_\mathrm{ir}^{(1)}}{4} \left( N_1^2 \Phi_1^2 + N_2^2 \Phi_2^2 + N_3^2 \Phi_3^2 \right) \nonumber \\ &  + \frac{\lambda_\mathrm{ir}^{(2)}}{4}\, N^2 \Phi^2 \,. \label{eq:F_ir}
\end{align}
In these expressions, $N^2 \equiv \sum _i N_i^2$ and $\Phi^2 \equiv \sum _i \Phi_i^2$. The quadratic Landau coefficients are defined in the standard way, $a_\mu = a_{\mu,0}(T-T_{\mu})$, with $\mu = \mathrm{r, i}$, the coefficient $a_{\mu,0} > 0$ and $T_{\mu}$ denote the bare transition temperatures. The Landau coefficients $u_{\mu}$ refer to quartic terms; $\gamma_{\mu}$, to trilinear terms; $\lambda_{\mu}$, to biquadratic terms; and $\kappa_{\mu}$, to quadrilinear terms. Note that, while the little group irreps have uniquely defined matrices, the space group irreps are arbitrary up to minus signs in certain off diagonal elements, which do not affect their characters. Similarly, shifting the origin by a unit cell changes the signs of certain components of the order parameters. As a result, the sign of the trilinear term is arbitrary, and is valid only for a fixed origin choice.

The fact that the bare iCDW and rCDW transition temperatures are different, $T_\mathrm{r}\neq T_\mathrm{i}$, is a consequence of the two order parameters transforming as different irreps. Nevertheless, the renormalization-group calculation of Ref.~\onlinecite{Park2021} found that for certain regimes of the interaction parameters, the two transition temperatures can be comparable. Even if this is not the case, a sufficiently strong trilinear coefficient $\gamma_\mathrm{ir}$ will generally cause the two transitions to happen simultaneously and in a first-order fashion, regardless of the sign of $\gamma_\mathrm{ir}$. 

The free energy in Eq.~\eqref{eq:Ftot} allows for various types of mixed phases. To gain insight into the typical global minima of this free energy, we note that it has the exact same functional form as the free energy for the coupled in-plane and out-of-plane rCDW orders studied by us in Ref.~\onlinecite{Christensen2021}. In that case, it was shown that much of the phase diagram is dominated by two phases in particular, which in our situation translate to a \QQ{} phase, where all three $N_i$ and all three $\Phi_i$ are non-zero, and to a \Q{} phase, where only $\Phi_{i}$, $\Phi_{j}$, and $N_{l}$ are non-zero, with $(i, j, l)$ denoting a permutation of $(1,2,3)$. Indeed, it is clear that these two mixed configurations are those that minimize the energy of the trilinear term with coefficient $\gamma_\mathrm{ir}$. Whether the \QQ{} or the \Q{} phase is realized depends on the quadrilinear and biquadratic coefficients, as discussed in Ref.~\onlinecite{Christensen2021}. 

Of course, it is also possible to realize non-mixed phases, such as a pure triple-$\mathbf{Q}$ rCDW phase or a pure single-$\mathbf{Q}$ iCDW phase. The latter is the only case in which iCDW order does not trigger rCDW order, as also discussed in Ref.~\onlinecite{Park2021}. Because these non-mixed phases cannot explain the experimental observations of time-reversal symmetry-breaking and a $2\times2$ increase of the unit cell, we will focus on the mixed iCDW-rCDW phases hereafter.

\begin{figure}
    \includegraphics[width=1\columnwidth]{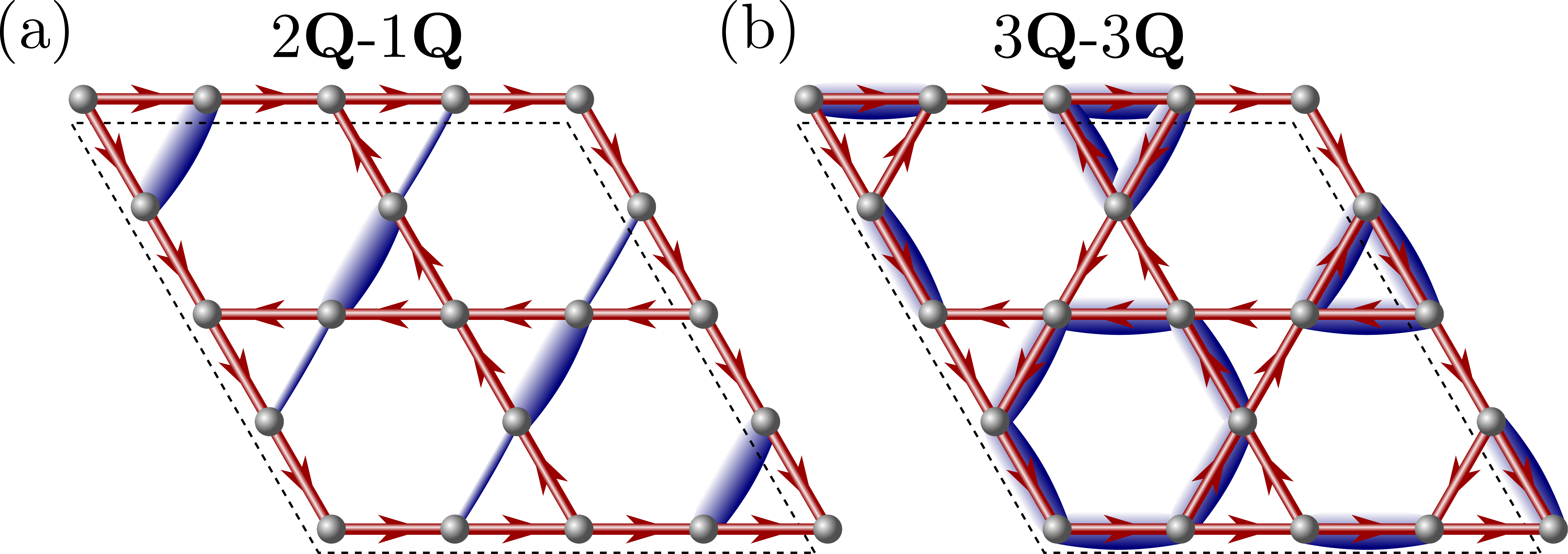}
    \caption{\label{fig:mixed_phases}Illustration of the two mixed iCDW-rCDW phases that can minimize the coupled free energy, Eq.~\eqref{eq:Ftot}. For concreteness, we show the case of intra-orbital iCDW order (i.e. $mM_2^+$) made out of the states from the same vHs. The rCDW order is also intra-orbital (i.e. $M_1^+$). (a) The \Q{} phase consists of a double-$\mbf{Q}$ iCDW phase combined with a single-$\mbf{Q}$ rCDW, such that the three wave-vectors are a permutation of $(\mathbf{Q}_{1}, \mathbf{Q}_{2}, \mathbf{Q}_{3})$. Blue bonds denote the bond distortions promoted by the rCDW order, whereas the red arrows denote the currents generated by the iCDW order. (b) The \QQ{} phase consists of a triple-$\mbf{Q}$ iCDW coexisting with a triple-$\mbf{Q}$ rCDW phase.}
\end{figure}

\begin{table*}
\centering
\begin{tabular}{C{0.14\textwidth}C{0.26\textwidth}C{0.26\textwidth}C{0.16\textwidth}C{0.14\textwidth}}
\hline 
\hline
\textbf{Mixed iCDW-rCDW} & \textbf{iCDW type} & \textbf{Subsidiary order} & \textbf{External field coupling} & \textbf{M.S.G.}\tabularnewline
\hline 
\hline 
\addlinespace
 & Any intra-orbital  & Ferromagnetic & $B_{z}$ & $P6/mm'm'$ \tabularnewline
 & $(mM_{2}^{+})$ & $(m\Gamma_{2}^{+},\,A_{2g})$ & & (\#191.240) \tabularnewline
\addlinespace
 & $p_{-}$-$p_{-}$ symm. inter-orbital  & Magnetic octupolar  & $(\mathbf{B}_{\parallel}\times\boldsymbol{\varepsilon}_{\parallel})\cdot\hat{\mathbf{z}}$ & $P6'/m'mm'$ \tabularnewline
 & $(mM_{3}^{+})$ & $(m\Gamma_{3}^{+},\,B_{2g})$ & & (\#191.239) \tabularnewline
\addlinespace
 & $p_{-}$-$p_{-}$ antisymm. inter-orbital & Magnetic octupolar  & $\mathbf{B}_{\parallel}\cdot\boldsymbol{\varepsilon}_{\parallel}$ & $P6'/m'm'm$ \tabularnewline
 & $(mM_{4}^{+})$ & $(m\Gamma_{4}^{+},\,B_{1g})$ & & (\#191.238) \tabularnewline
\addlinespace
$3\mathbf{Q}$-$3\mathbf{Q}$ & $p_{-}$-$m$ symm. inter-orbital & Magnetic monopolar & $E_{z}B_{z}$, $\mathbf{E}_{\parallel}\cdot\mathbf{B}_{\parallel}$ & $P6'/m'm'm'$ \tabularnewline
$\left(\mu\sim\Phi^{3}\right)$ & $(mM_{1}^{-})$ & $(m\Gamma_{1}^{-},\,A_{1u})$ & & (\#191.241) \tabularnewline
\addlinespace
 & $p_{-}$-$m$ antisymm. inter-orbital  & Magnetic toroidal dipolar  & $(\mathbf{E}\times\mathbf{B})\cdot\hat{\mathbf{z}}$ & $P6/m'mm$ \tabularnewline
 & $(mM_{2}^{-})$ & $(m\Gamma_{2}^{-},\,A_{2u})$ & & (\#191.235)\tabularnewline
 \addlinespace
 & $p_{+}$-$m$ symm. inter-orbital & Magnetic toroidal octupolar & $E_{z}\left(\mathbf{B}_{\parallel}\cdot\boldsymbol{\varepsilon}_{\parallel}\right)$ & $P6'/mm'm$ \tabularnewline
 & $(mM_{3}^{-})$ & $(m\Gamma_{3}^{-},\,B_{2u})$ & &(\#191.236) \tabularnewline
\addlinespace
 & $p_{+}$-$m$ antisymm. inter-orbital & Magnetic toroidal octupolar & $\mathbf{E}\cdot(\mathbf{B}_{\parallel}\times\boldsymbol{\varepsilon}_{\parallel})$ &$P6'/mmm'$\tabularnewline
 & $(mM_{4}^{-})$ & $(m\Gamma_{4}^{-},\,B_{1u})$ & &(\#191.237)\tabularnewline
\addlinespace
\hline 
\addlinespace
$2\mathbf{Q}$-$1\mathbf{Q}$  & $mM_2^+$, $mM_3^-$, $mM_4^-$ & Orthorhombic distortion & $\boldsymbol{\varepsilon}_{\parallel}$ & $C_{a}mmm$ \tabularnewline
$\left(\eta\sim\Phi^{2}\right)$ &  & $(m\Gamma_{5}^{+},\,E_{2g})$ & & (\#65.489) \tabularnewline
\addlinespace
& $mM_3^+$, $mM_4^+$, $mM_1^-$, $mM_2^-$ & Orthorhombic distortion & $\boldsymbol{\varepsilon}_{\parallel}$ & $C_{a}mma$ \tabularnewline
&  & $(m\Gamma_{5}^{+},\,E_{2g})$ & & (\#67.509) \tabularnewline
\addlinespace
\hline
\hline
\end{tabular}
\caption{\label{tab:summary_table}Summary of the different types of mixed iCDW-rCDW configurations found in this work, arising from the low-energy electronic states associated with the two occupied vHs ($p_-$-type) and the two unoccupied vHs ($p_+$-type and $m$-type) shown in Fig.~\ref{fig:orbitals_and_bands}(c). Note that, in rows 4 and 5, $p_-$ refers only to the occupied vHs closest to the Fermi level. In all cases, the rCDW is described by an $M_1^+$ order parameter, whereas the iCDW order parameter $\boldsymbol{\Phi}$ is given by the second column of the table. There are two different kinds of mixed state, \QQ{} and \Q{}. The first displays different types of subsidiary uniform magnetic order, whose order parameter $\mu$ scales as $\Phi^3$, whereas the latter displays only an orthorhombic distortion $\eta$ that scales quadratically with $\Phi$. For each subsidiary order we include both the space group irreps (using the convention of Ref.~\onlinecite{Bradley1972Mathematical}) and point group irreps (using the convention of Ref.~\onlinecite{Cracknell1979Kronecker}; in the first seven rows, the irreps are odd under time-reversal whereas in the last two rows, it is even). The fourth column shows which combination of external fields couples to the subsidiary order parameter $\mu$ or $\boldsymbol{\eta}$. Here, $\mathbf{B}$ is the magnetic field, $\mathbf{E}$ is the electric field, and $\boldsymbol{\varepsilon}_{\parallel}$ is the in-plane strain field given by Eq. (\ref{eq:strain}). The last column shows the name and number of the magnetic space groups (M.S.G.) of each mixed state in the Belov-Neronova-Smirnova notation. Primed operations refer to point symmetry operations followed by time-reversal. For example, $6'$ denotes a 60$^\circ$ rotation followed by time-reversal, and $m'$ denotes a mirror reflection followed by time reversal.}
\end{table*}

Figure~\ref{fig:mixed_phases} illustrates both types of mixed rCDW-iCDW phases, \QQ{} and \Q{}, for the particular case of an intra-orbital $mM_2^+$ iCDW order parameter. As explained above, we are only considering the intra-orbital rCDW order, as it transforms as $M_1^+$. As expected, the kagome lattice displays not only bond distortions but also loop currents. We emphasize that a pure double-$\mbf{Q}$ iCDW phase or a pure triple-$\mbf{Q}$ iCDW phase are not minima of the free energy, as they can only arise in conjunction with rCDW order. Moreover, for the \QQ{} phase, while the sign of the product $\Phi_1 \Phi_2 \Phi_3$ does not change the loop-currents configuration~\cite{Park2021}, the sign of $N_1 N_2 N_3$ distinguishes between a star-of-David and a tri-hexagonal bond-order configuration~\cite{Christensen2021}.

\section{Experimental signatures of the mixed CDW states: occupied vHs}\label{sec:implications}

By focusing on the low-energy electronic states near the two vHs below the Fermi level at the M point, we found three possible types of iCDW states, illustrated in Fig.~\ref{fig:iCDW_phases}: intra-orbital ($mM_2^+$), symmetric inter-orbital ($mM_3^+$), and antisymmetric inter-orbital ($mM_4^+$). When coupled to the intra-orbital rCDW order parameter ($M_1^+$), they in turn give rise to two different types of mixed iCDW-rCDW states, denoted \QQ{} and \Q{}, illustrated in Fig.~\ref{fig:mixed_phases} in the case of intra-orbital iCDW order. There are therefore six different candidate CDW states formed out of the occupied vHs that could explain the experimental observations of a $2\times2$ unit cell increase and time-reversal symmetry-breaking below $T_{\mathrm{CDW}}$. In this section, we discuss the experimental signatures of these 6 states, focusing on the magnetic and structural properties that can be probed experimentally to unambiguously distinguish between them. A summary of the results of this section and of Sec.~\ref{sec:unoccupied} is contained in Table~\ref{tab:summary_table} for convenience.

\subsection{Finite-momentum magnetism: spin-density wave}

\begin{figure*}
    \includegraphics[width=\textwidth]{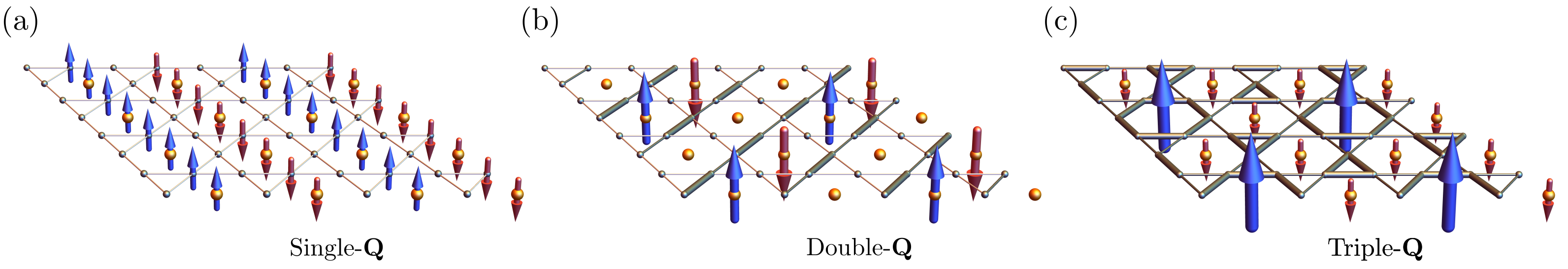}
    \caption{Illustrations of the spin density-wave patterns induced by intra-orbital iCDW in the case of single-$\mathbf{Q}$ [panel(a)], double-$\mathbf{Q}$ [panel(b)], and triple-$\mathbf{Q}$ [panel(c)] order. Note that the spin density-wave has peaks at the Sb atoms, rather than at the V atoms. In panels (b) and (c), corresponding to the \Q{} and \QQ{} states, respectively, rCDW order must accompany the iCDW one. The bond-distortion patterns corresponding to these rCDW orders are illustrated as well. In the \QQ{} case, the local environment experienced by the sites with spin-up is different from that experienced by the sites with spin-down, resulting in a net magnetic moment. \label{fig:rSDW_mm2+}}
\end{figure*}

In the presence of spin-orbit coupling, the iCDW orders must necessarily induce spin-density wave (SDW) orders~\cite{Klug2018}. Indeed, an iCDW corresponds to an imaginary hopping between two sites on the lattice. This leads to ordering of the electrons' orbital angular momentum, which in turn results in ordering of the spin angular momentum in the presence of spin-orbit coupling. 

In terms of symmetry, we can define other fermionic bilinears built out of the Bloch states of the saddle points that transform as the same irreps as the iCDW order parameters. For instance, consider the intra-orbital SDW order parameter with wave-vector $\mathbf{Q}_{1}$:
\begin{equation}
        \vec{\Delta}_{1} = \sum_{\mbf{k}\sigma \sigma'} \langle
        c_{\mbf{k}+\mbf{Q}_{2},\sigma}^\dagger \vec{\tau}_{\sigma \sigma'} c_{\mbf{k}+\mbf{Q}_{3},\sigma'}^{\phantom{\dagger}} + \text{H.c.}
        \rangle\,, \label{eq:SDW} 
\end{equation}
where $\tau^{i}$ denotes a Pauli matrix in spin-space. Using the local coordinate system of V$_1$ to label the spin directions, it turns out that $\Delta_1^z$ transforms as $mM_2^+$, $\Delta_1^x$ transforms as $mM_3^+$, and $\Delta_1^y$, as $mM_4^+$. However, the directions of the in-plane spins for the $mM_3^+$ and $mM_4^+$ order parameters are different for different wave-vectors. As a result, labeling these order parameters with a Cartesian direction could be misleading. Instead, we note that the $mM_3^+$ order parameter gives rise to a spin stripe with spins normal to the wave-vector, whereas for $mM_4^+$ the spins are parallel to the wave-vector. As a result, we refer to the $mM_3^+$ and $mM_4^+$ SDW orders as $\Delta_i^\perp$ and $\Delta_i^\parallel$ respectively, instead of $\Delta_i^x$ and $\Delta_i^y$.

Since there are SDW order parameters that transform as the same irrep as the iCDW order parameters, the condensation of the iCDW order parameter $\Phi_1$ leads to the condensation of an intra-orbital SDW at the same wave-vector, since it is bilinearly coupled in the free-energy to one of the components of $\vec{\Delta}$. In particular, the magnetic moments in the SDW phase are polarized along different axes depending on the nature of the iCDW state: $z$-axis, for intra-orbital iCDW; the axis normal to the wave-vector, for symmetric inter-orbital iCDW; and the axis parallel to the wave-vector, for antisymmetric inter-orbital iCDW. 

\begin{figure*}
\includegraphics[width=\textwidth]{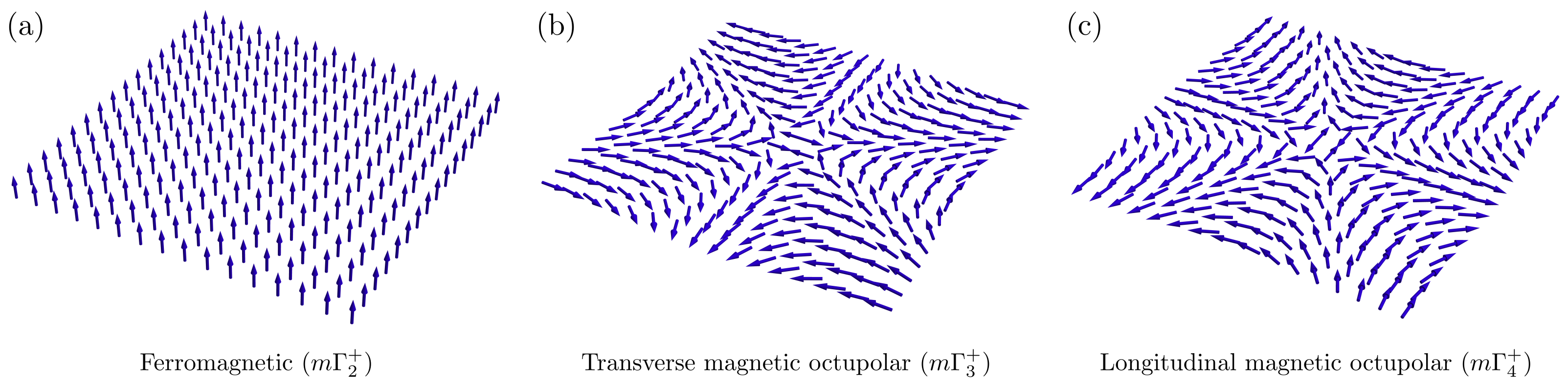}
\caption{Illustrations of different types of uniform magnetic order: (a), ferromagnetism; (b) transverse magnetic octupolar order; (c) longitudinal magnetic octupolar order. These subsidiary orders appear in the \QQ{} states with intra-band, symmetric inter-band, and anti-symmetric inter-band iCDW orders generated from the occupied vHs, respectively.\label{fig:induced_mag_orders}}
\end{figure*}

This result provides another route to probe the existence of an iCDW, as neutron scattering experiments could in principle directly assess the existence of SDW order from the magnetic Bragg peaks that it creates. Of course, the feasibility of such a measurement will depend on the size of the SDW magnetic moment. Interestingly, polarized neutron scattering could further distinguish between the three different types of iCDW by determining the direction of the moments. One important caveat related to this last point is that inter-orbital SDW can also be induced, but with a moment direction that is generally different from that of the order parameter in Eq.~\eqref{eq:SDW}.

Figure~\ref{fig:rSDW_mm2+} illustrates the magnetic patterns of the SDW phases induced by single-$\mathbf{Q}$ (a), double-$\mathbf{Q}$ (b), and triple-$\mathbf{Q}$ (c) intra-band iCDW order. Rather than determining these magnetic configurations indirectly from the loop-current patterns, here we directly derive them from the symmetry properties of the relevant irrep -- which in this case is $mM_2^+$. As expected, a single-$\mathbf{Q}$ iCDW generates a stripe SDW with the moments oriented out-of-plane. Interestingly, in both the double- and triple-$\mathbf{Q}$ phases, the superposition of the different stripe SDWs results in peaks of the magnetization density at the Sb sites, rather than the V sites.

\subsection{Uniform magnetism: ferromagnetism and magnetic octupolar order}

In addition to the SDW orders that accompany the onset of iCDW orders, different types of subsidiary \emph{uniform} magnetic order also appear in the mixed iCDW-rCDW phases. They are described by the scalar order parameter
\begin{equation}
    \mu = \Phi_1 \Phi_2 \Phi_3 \propto \mbf{N}\cdot\boldsymbol{\Phi} \,, \label{eq:mu_moment} 
\end{equation}
which is clearly odd under time-reversal and has zero wave-vector. Note that $\mu$ is only non-zero in the \QQ{} mixed phase, and the \Q{} phase does not display subsidiary uniform magnetic order.

Importantly, the transformation properties of $\mu$, and thus the nature of the uniform magnetic order, depend crucially on the type of iCDW order. For intra-orbital iCDW order ($mM_2^+$), $\mu$ transforms as the time-reversal-odd $A_{2g}$ irrep of the $D_{6h}$ point group (or, equivalently, $m\Gamma_2^{+}$ of $P6/mmm$ using the notation of INVARIANTS, which follows the Cracknell-Davies-Miller-Love tables~\cite{Cracknell1979Kronecker, Aroyo2006}). Therefore, $\mu$ corresponds to ferromagnetic order with moments pointing out-of-plane, as illustrated in Fig.~\ref{fig:induced_mag_orders}(a). This is in agreement with the SDW pattern shown in Fig.~\ref{fig:rSDW_mm2+}(c): in a single unit cell there are smaller moments at three Sb sites pointing in one direction and a bigger moment at another Sb site pointing in the opposite direction. Because the local environments experienced by the two types of Sb sites are inequivalent due to the accompanying bond-distortion pattern, the four moments do not cancel, and a net dipolar magnetic moment is generated. Alternatively, one can conclude that a net magnetic moment must be present because the application of time-reversal symmetry cannot be undone by a translation. We note that Ref.~\onlinecite{Park2021} previously pointed out that a triple-$\mathbf{Q}$ iCDW order induces an out-of-plane magnetization. Indeed, the magnetic space group of the \QQ{} configuration with intra-band iCDW order, shown in the last column of Table~\ref{tab:summary_table}, allows a non-zero out-of-plane magnetic dipole moment. 

From the definition of $\mu$, we conclude that the induced magnetization scales as $|\boldsymbol{\Phi}|^3$, where $|\boldsymbol{\Phi}|$ is the magnitude of the iCDW order parameter. Thus, because the induced SDW $\Delta^z$ scales linearly with $|\boldsymbol{\Phi}|$, there is a well-defined relationship between the uniform magnetization and the finite-momentum magnetization, $\mu \sim (\Delta^z)^3$. This result provides yet another route to probe iCDW order in \AVS{} through a coupling to the SDW order.

In the case of inter-orbital iCDW orders, a finite $\mu$ implies different types of uniform magnetic orders, as it transforms as the time-reversal odd $B_{2g}$ and $B_{1g}$ irreps of $D_{6h}$ for the symmetric ($mM_3^+$) and anti-symmetric ($mM_4^+$) types of iCDW order, respectively. In terms of the irreps of the space group $P6/mmm$, they correspond to $m\Gamma_3^+$ and $m\Gamma_4^+$, respectively. Physically, an order parameter with these transformation properties results in \emph{magnetic octupolar order} \cite{Hayami2018}. 

The real-space configuration of magnetic moments associated with these two types of magnetic octupolar order are shown in Figs.~\ref{fig:induced_mag_orders}(b)-(c), which we dub transverse [$B_{2g}$, panel(b)] and longitudinal [$B_{1g}$, panel (c)] ferro-octupolar orders. In both cases, the magnetic moments point in-plane, but are subjected to different spatial modulations. Neither configuration results in a net magnetic moment, in agreement with the fact that, for the triple-$\mathbf{Q}$ inter-band iCDW orders, the loop currents above and below the kagome plane cancel each other, as shown in Fig.~\ref{fig:iCDW_phases}.

The most straightforward way to probe these subsidiary magnetic octupolar orders is to assess their magneto-striction properties. We start by constructing the following vectors from the in-plane components of the magnetic field and of the strain tensor
\begin{equation}
   \mathbf{B}_{\parallel} = \begin{pmatrix}
    B_x \\
    B_y
    \end{pmatrix}\,, \qquad
    \boldsymbol{\varepsilon}_{\parallel} = \begin{pmatrix}
    \varepsilon_{x^2-y^2} \\
    -\varepsilon_{xy}
    \end{pmatrix}\,. \label{eq:strain}
\end{equation}
Here, $\varepsilon_{ij} \equiv (\partial_i u_j + \partial_j u_i )/2$, where $\mathbf{u}$ denotes the lattice displacement vector. In order to be consistent with the irrep matrices of Ref.~\onlinecite{StokesInvariants}, we choose a global coordinate axis with $\hat{\mathbf{y}}$ parallel to the crystallographic $a$ axis (the [100] direction).
Using the fact that $\mathbf{B}_{\parallel}$ transforms as the $m\Gamma_6^+$ irrep, whereas $\boldsymbol{\varepsilon}_{\parallel}$ transforms as $\Gamma_5^+$, we derive the following magneto-striction free-energy term in the \QQ{} state with symmetric inter-orbital iCDW order:
\begin{equation}
    \mathcal{F}^{(mM_3^+)}_{\rm mag-str} \sim \mu \left( \mathbf{B}_{\parallel} \times \boldsymbol{\varepsilon}_{\parallel} \right) \cdot \hat{\mathbf{z}} \,, \label{eq:magstrsym}
\end{equation}
where $\mu$ is the magnetic octupolar moment defined in Eq.~\eqref{eq:mu_moment}. Conversely, for the \QQ{} state with antisymmetric inter-orbital iCDW order, we find:
\begin{equation}
    \mathcal{F}^{(mM_4^+)}_{\rm mag-str} \sim \mu \left( \mathbf{B}_{\parallel} \cdot \boldsymbol{\varepsilon}_{\parallel} \right)\,.
\end{equation}
The first of these equations, Eq.~\eqref{eq:magstrsym}, implies that, in the \QQ{} state with symmetric inter-orbital iCDW order, application of an in-plane magnetic field generates strain in the ``transverse" direction (in the abstract subspace of two-dimensional irreps). For example, a magnetic field applied along the $y$-axis (which is the crystallographic $a$-axis) will generate the strain component $\epsilon_{x^2 - y^2}$. 
In contrast, in the case of antisymmetric inter-orbital iCDW order, the strain generated by an in-plane magnetic field is ``longitudinal" -- e.g., a finite $B_y$ induces $\epsilon_{xy}$. Of course, the reverse is also true: shear strain $\epsilon_{xy}$ induces magnetic moments along the $y$-axis in the latter case and along the $x$-axis in the former. These unique magneto-striction properties could be used to experimentally probe the character of the \QQ{} phases. A summary of these results is contained in Table~\ref{tab:summary_table}.

\subsection{Lattice distortion: threefold rotational symmetry-breaking}

While the \QQ{} states display uniform magnetic order, the \Q{} states are accompanied by a uniform lattice distortion that breaks the threefold rotational symmetry of the kagome lattice. This is apparent in Fig.~\ref{fig:mixed_phases}(a), as the loop-current and bond-distortion patterns are not invariant under  $120^\circ$ rotation. Formally, we can construct an order parameter that is quadratic in the iCDW order parameters and transforms as the same $\Gamma_5^+$ irrep as the in-plane strain $\boldsymbol{\varepsilon}_{\parallel}$ defined in Eq.~(\ref{eq:strain}):
\begin{equation}
    \boldsymbol{\eta}_{1} = \begin{pmatrix}
        \Phi_1^2 + \Phi_3^2 - 2 \Phi_2^2 \\
        \sqrt{3}(\Phi_3^2 - \Phi_1^2)
    \end{pmatrix}\,. \label{eq:eta1}
\end{equation}
Clearly, $\boldsymbol{\eta}_{1}$ is only nonzero in the \Q{} phase, leading to an orthorhombic distortion of the lattice. Similarly, we can define an order parameter that also transforms as $\Gamma_5^+$ and that depends explicitly on the rCDW order parameter $\mathbf{N}$:
\begin{equation}
    \boldsymbol{\eta}_{2} = \begin{pmatrix}
        N_1 \Phi_2 \Phi_3 + \Phi_1 \Phi_2 N_3 - 2 \Phi_1 N_2 \Phi_3 \\
        \sqrt{3}(\Phi_1 \Phi_2 N_3 - N_1 \Phi_2 \Phi_3)
    \end{pmatrix} \,. \label{eq:eta2}
\end{equation}
Note that, in contrast to $\boldsymbol{\eta}_{1}$, $\boldsymbol{\eta}_{2}$ is quartic in $\boldsymbol{\Phi}$.

Recent experiments have reported the breaking of the threefold rotational symmetry of the kagome lattice inside the CDW phase of the \AVS{} compounds ~\cite{Xiang2021,Zhao2021Cascade,Li2022Rotation,Nie2022Charge,Xu_MOKE,Guo2022Field-tuned}. In contrast to time-reversal symmetry-breaking, however, the rotational symmetry seems to be broken only well below $T_{\mathrm{CDW}}$. Therefore, this type of order does not onset before the CDW and breaks translational symmetry; as such, it should not be classified as a nematic phase. 

Previously, the breaking of threefold rotational symmetry was proposed to arise from an admixture between two real CDWs, one with and one without out-of-plane modulation \cite{Park2021,Ratcliff2021,Christensen2021}. Our analysis shows that it could also be explained by an admixture between real and imaginary CDWs in the \Q{} state, without the need to invoke an out-of-plane component of the wave-vector. In this respect, it is tempting to attribute the experimental observations to a transition from the \QQ{} phase to the \Q{} phase. However, in several of these experiments, the threefold rotational symmetry breaking was observed in the presence of an in-plane magnetic field, which presumably aligns the different orthorhombic domains. As we showed in the previous subsection, $\mathbf{B}_{\parallel}$ induces a finite orthorhombic distortion $\boldsymbol{\varepsilon}_{\parallel}$ in the \QQ{} phase with inter-orbital iCDW order. Therefore, to shed light on this issue, it is important that future experiments clarify the relationship between the in-plane field and the threefold rotational symmetry breaking. 

\subsection{Impact of out-of-plane modulation}
Our analysis so far has focused solely on iCDW and rCDW orders with in-plane wave-vectors $\mbf{Q}_{i}$ coinciding with the M$_i$ points of the BZ. X-ray experiments on \AVS, however, reveal that the charge-order wave-vector can also have a finite commensurate $z$ component, resulting in a $2\times2\times2$ or $2\times2\times4$ unit cell in the ordered state \cite{Ortiz2021Fermi,Li2021Observation}. The precise $c$-axis modulation depends not only on the alkali-metal atom $A$, but also on temperature and pressure \cite{Stahl2021Temperature,Miao2022,Xiao2022coexistence}.

It is straightforward to extend our analysis to the case of an ordering wave-vector along the $\rm{M}-\rm{L}$ line of the BZ, which we label by its $z$-axis component $Q^z$. Based on the experimental results mentioned above, we restrict our analysis to two different possible out-of-plane modulations for the iCDW order parameter, $Q^z_{\mathrm{iCDW
}} = 1/2$ and $Q^z_{\mathrm{iCDW
}} = 1/4$ (in reciprocal lattice vector units). As we saw in Sec.~\ref{sec:free_energy}, the intrinsic coupling between the iCDW and rCDW order parameters arises from the trilinear term in the free energy of Eq. (\ref{eq:F_ir}). In order for that term to be preserved, it must follow that
\begin{equation}
  Q^z_{\mathrm{rCDW}} = 2\,Q^z_{\mathrm{iCDW}} \, ,
\end{equation}
i.e. the rCDW modulation must be equal to half of the iCDW modulation. By symmetry, it also follows that the induced SDW has the same modulation as the iCDW, $Q^z_{\mathrm{SDW}} = Q^z_{\mathrm{iCDW}}$.

The properties of the subsidiary order parameters $\mu$ and $\boldsymbol{\eta}$ can also be determined in a straightforward way. Because $\mu$ is cubic in $\Phi_i$, it has the same $z$-component of the wave-vector as the iCDW order parameter, $Q^z_{\mu} = Q^z_{\mathrm{iCDW}}$ (recall that we are restricting our analysis to $Q^z_{\mathrm{iCDW}} = 1/2, 1/4$). Consequently, in the \QQ{} state, the induced magnetic order is uniform in the kagome planes but exhibits the same modulation along the $c$-axis as the iCDW order parameter. This implies that antiferromagnetic order emerges in the case of intra-orbital iCDW; for inter-orbital iCDW, the same antiferromagnetic pattern arises if uniaxial in-plane stress is applied.

As for $\boldsymbol{\eta}$, which is proportional to the square of $\Phi$, its wave-vector remains at the $\Gamma$ point, i.e. $Q^z_{\boldsymbol{\eta}}=0$. Consequently, the \Q{} state displays an orthorhombic distortion regardless of the modulation of the iCDW. To arrive at this conclusion, it is important to note that, unlike the $Q^z_{\mathrm{iCDW}} = 0, 1/2$ cases, the iCDW order parameter becomes a \emph{complex-valued} three-component order parameter for $Q^z_{\mathrm{iCDW}} = 1/4$, since $\mathbf{Q}_{\mathrm{iCDW}} \neq -\mathbf{Q}_{\mathrm{iCDW}}$. As a result, $\boldsymbol{\eta}_1$ in Eq. (\ref{eq:eta1}) must be rewritten with $\Phi_i^2 \rightarrow |\Phi_i|^2$; in contrast, $\boldsymbol{\eta}_2$ in Eq. (\ref{eq:eta2}) remains the same.

\section{Candidate charge orders: unoccupied van Hove singularities} \label{sec:unoccupied}

We now repeat the analyses of Secs.~\ref{sec:occupied}, \ref{sec:free_energy}, and \ref{sec:implications} for the iCDW phases that arise from the Bloch states associated with the unoccupied vHs of Fig.~\ref{fig:orbitals_and_bands}(c). We recall that the unoccupied vHs closest to the Fermi level transforms as the $M_3^+$ irrep and is of $p$-type, whereas the second unoccupied vHs closest to the Fermi level is of $m$-type and transforms as the $M_2^-$ irrep. Since the steps are the same as in the previous sections, we do not delve into details of the calculation and just present the main results.

\subsection{iCDW phases from the unoccupied vHs only}

Mirroring what we did for the two occupied vHs, we first consider the three different types of iCDW order that emerge from combinations of electronic states associated with the two unoccupied vHs. As shown in the insets of Fig.~\ref{fig:orbitals_and_bands}(c), the local $d_{yz}$ orbitals contribute to the $M_3^+$ vHs whereas the local $d_{xz}$ orbitals contribute to the $M_2^-$ vHs. For this reason, we continue using the nomenclature of intra-orbital and inter-orbital iCDW.

Labeling the electronic states near the $M_3^+$ vHs by the operator $c^\dagger_\mbf{k\sigma}$ and the electronic states near the $M_2^-$ vHs by $d^{\dagger}_\mbf{k\sigma}$, we obtain the iCDW order parameters as in Table \ref{tab:iCDW_order_parameters}. While the intra-orbital $\Phi_i^c$ and $\Phi_i^d$ iCDW order parameters still transform as $mM_2^+$, the inter-orbital $\Phi_i^s$ and $\Phi_i^a$ now transform as $mM_3^-$ and $mM_4^-$, respectively. Remarkably, they couple to the $M_1^+$ rCDW order parameter $N_i$ in the same way as their counterparts in Table \ref{tab:iCDW_order_parameters}, i.e. the coupled iCDW-rCDW Landau free-energy acquires the same form as Eq.~(\ref{eq:Ftot}). Consequently, there are two viable mixed iCDW-rCDW states, the \QQ{} and the \Q{} ones.

The \QQ{} mixed configuration displays subsidiary uniform magnetic order described by the cubic order parameter $\mu \sim \Phi^3$ defined in Eq.~\eqref{eq:mu_moment} above. Like the magnetic octupolar orders generated in the case of inter-band iCDW order involving the occupied vHs, these subsidiary magnetic orders do not have a net magnetic dipole moment. However, they are odd under inversion, as indicated by their transformation properties: $m\Gamma_3^-$ (or time-reversal odd $B_{2u}$), for the case of the symmetric inter-orbital iCDW, and $m\Gamma_4^-$ (or time-reversal odd $B_{1u}$), for the anti-symmetric inter-orbital iCDW. Therefore, they can be identified as two different types of \emph{magnetic toroidal octupolar} order, following the classification of Ref.~\onlinecite{Hayami2018}. Indeed, the key property of an octupolar magnetic toroidal moment is that it changes sign not only under time reversal, but also under spatial inversion. These magnetic toroidal octupolar orders can be probed experimentally by a combination of magnetic field $\mathbf{B}$, in-plane uniaxial strain $\boldsymbol{\varepsilon}_{\parallel}$, and electric field $\mathbf{E}$. Using group theory, we find the following free-energy couplings between the external fields and the magnetic toroidal octupolar moment $\mu$, corresponding to magneto-electric-striction terms:
\begin{equation}
    \mathcal{F}^{(mM_3^-)}_{\rm mag-el-str} \sim \mu \, E_{z}\left(\mathbf{B}_{\parallel}\cdot\boldsymbol{\varepsilon}_{\parallel}\right) \,,
\end{equation}
as well as
\begin{equation}
    \mathcal{F}^{(mM_4^-)}_{\rm mag-el-str} \sim \mu \, \mathbf{E}\cdot(\mathbf{B}_{\parallel}\times\boldsymbol{\varepsilon}_{\parallel}) \,.
\end{equation}
Therefore, application of in-plane uniaxial strain induces \emph{multiferroic} order characterized by an in-plane magnetic moment (whose direction can be longitudinal or transverse to the strain direction) and an out-of-plane electric polarization.

As for the \Q{} mixed configuration, its subsidiary uniform order is not magnetic, but orthorhombic, like in the case of inter-orbital iCDW made out of the occupied vHs. It is described by the same order parameter $\boldsymbol{\eta}$ defined in Eq.~(\ref{eq:eta1}), which scales as the square of the iCDW order parameter $\Phi$. Table~\ref{tab:summary_table} summarizes the properties of the mixed iCDW-rCDW states generated by the unoccupied vHs, including the magnetic space groups that describe them.

\subsection{iCDW phases from mixed occupied and unoccupied vHs}

So far we have considered the two pairs of vHs -- occupied and unoccupied -- separately. The reasoning is that the energy separation between these pairs, of the order of $200$ meV [see Fig. \ref{fig:orbitals_and_bands}(c)], is significant enough that it is likely to prevent instabilities driven by the coupling between states associated with occupied and unoccupied vHs. Nevertheless, it is possible that this energy splitting could be reduced by correlations not captured by DFT, or by different doping schemes. Moreover, even within DFT, upon moving away from the M point along the $\rm{M}-\rm{L}$ line, the energy splitting between the pairs of vHs changes as a function of out-of-plane momentum \cite{Kang2022Twofold}. 

Therefore, and for the sake of completeness, we briefly discuss the properties of the possible iCDW order parameters obtained by combining states from vHs above and below the Fermi level. The only combinations that give iCDW orders not already discussed above are the symmetric and anti-symmetric inter-orbital iCDW order parameters made out of states from the $m$-type vHs and from the $p_-$-type vHs that is closest to the Fermi level. Specifically, the corresponding order parameters $\Phi_i^s$ and $\Phi_i^a$, as defined in Table \ref{tab:iCDW_order_parameters}, transform as the irreps $mM_1^-$ and $mM_2^-$. The corresponding coupled iCDW-rCDW free-energy expansions, with the rCDW order parameter $N_i$ transforming as $M_1^+$, are once again identical to that written in Eq.~(\ref{eq:Ftot}), which promotes the mixed \QQ{} and \Q{} configurations.

Similar to the cases studied above, the \Q{} configuration breaks the threefold rotational symmetry of the kagome lattice by causing an orthorhombic distortion given by the order parameter $\boldsymbol{\eta}$ [see Eq.~(\ref{eq:eta1})]. On the other hand, the \QQ{} mixed configuration is accompanied by uniform magnetic order, whose order parameter $\mu$ transforms as either $mM_1^-$ (i.e. time-reversal-odd $A_{1u}$), for symmetric inter-orbital iCDW order, or as $mM_2^-$ (i.e. time-reversal-odd $A_{2u}$), for anti-symmetric inter-orbital iCDW order. While the former corresponds to \emph{magnetic monopolar} order, the latter is a \emph{magnetic toroidal dipolar} order~\cite{Hayami2018}. They have unique magneto-electric properties, as described by the following Landau free-energy couplings:
\begin{equation}
    \mathcal{F}^{(mM_1^-)}_{\rm mag-el} \sim \mu \, E_{z} B_z \,,
\end{equation}
and
\begin{equation}
    \mathcal{F}^{(mM_2^-)}_{\rm mag-el} \sim \mu \, (\mathbf{E} \times \mathbf{B})\cdot \hat{\mathbf{z}} \,.
\end{equation}
We note that magnetic toroidal dipolar order has been previously proposed to explain the time-reversal symmetry-breaking state of the \AVS{} kagome compounds in Ref.~\onlinecite{Lovesey2022hidden}. Indeed, the resulting magnetic space group name associated with the \QQ{} configuration with $mM_2^-$ iCDW order (see Table \ref{tab:summary_table}) is the same as that discussed in Ref.~\onlinecite{Lovesey2022hidden}. The main difference is that, in our case, the toroidal moment is a subsidiary order of a triple-$\mathbf{Q}$ iCDW order that also breaks the translational symmetry of the crystal.

\section{Conclusions}\label{sec:conclusions}

We have combined phenomenology, DFT calculations, and group theory to derive the possible mixed iCDW-rCDW states of \AVS{} compounds that can arise from interactions involving the two pairs of occupied and unoccupied vHs closest to the Fermi level. Because these four vHs have different orbital and sublattice structures, as shown in Fig.~\ref{fig:orbitals_and_bands}, various types of rCDW and iCDW states are possible. Current experimental and first-principles results constrain the rCDW to transform as the $M_1^+$ irrep, which corresponds to intra-orbital/intra-vHs order in our model. On the other hand, the available experimental data does not allow one to unambiguously identify the type of iCDW order that is possibly realized in \AVS. Our findings, which are summarized in Table \ref{tab:summary_table}, reveal seven different possible iCDW order parameters involving these two pairs of vHs. As we discussed here, any of these iCDW configurations must be accompanied by an SDW pattern that shares the same symmetry properties; in the particular case of intra-orbital/intra-vHs iCDW, the magnetization density peaks at the Sb atoms, as shown in Fig. \ref{fig:rSDW_mm2+}. We note that one of such iCDW orders ($mM_2^+$) was previously identified in Ref.~\onlinecite{Park2021}, and that an alternative classification scheme was also put forward in Ref.~\onlinecite{Feng2021Low-energy}.

One of our main results, derived from the analysis of the coupled iCDW-rCDW free-energy expansion, is that, unless the iCDW order is unidirectional, it will be accompanied by rCDW order, either in a \QQ{} or in a \Q{} mixed configuration. The intrinsic coupling between these two types of order was previously discussed in Refs.~\onlinecite{Park2021,Lin2021}; here, we focused on the most favored minima of the full free-energy in the large parameter space available. 

Thus, because we identify seven types of iCDW order and two possible minima of the coupled free energy with the $M_1^+$ rCDW order, there are 14 different viable mixed iCDW-rCDW configurations, as shown in Table~\ref{tab:summary_table}. Although our list of seven iCDW orders was derived from the symmetry properties of the Bloch states at the four vHs, it is quite comprehensive, as there are only eight possible irreps for the iCDW order parameter with wave-vector corresponding to the M point. The only irrep that did not appear in our analysis is $mM_1^+$. The reason for this absence is because this irrep is trivial in the sense that it breaks no symmetries other than the translational symmetry (due to its finite wave-vector) and time-reversal symmetry. Thus, there is no combination of currents normal to the $c$ axis that gives rise to an order parameter that transform as $mM_1^+$. 

Our second main result is the identification of the experimental manifestations of these 14 different states. The formal assignment in terms of magnetic space groups is presented in the last column of Table~\ref{tab:summary_table}. While the seven \QQ{} phases have seven different magnetic space groups, the seven \Q{} configurations group in just two different magnetic space groups. In principle, detailed x-ray and neutron scattering experiments may be able to distinguish between these magnetic space groups and thus identify which iCDW order is realized in \AVS.

Conversely, these iCDW-rCDW mixed configurations display subsidiary uniform (i.e. zero wave-vector) orders that can be identified experimentally, as discussed in Table~\ref{tab:summary_table}. In particular, while the \Q{} phases have no uniform magnetic order, they all display an orthorhombic distortion that scales as the square of the iCDW order parameter, regardless of the type of iCDW order present. In contrast, each of the seven \QQ{} phases display different types of uniform magnetic order, whose order parameters scale as the cube of the iCDW order parameter. The simplest type of uniform magnetic order -- ferromagnetism -- is realized in the case of intra-orbital/intra-vHs iCDW order. In all cases that involve inter-vHs order, a more exotic type of uniform magnetism arises, namely, magnetic octupolar, magnetic toroidal, and magnetic monopolar order. Each of them displays unique magneto-striction, magneto-electric, or magneto-electric-striction properties, which can be probed by the appropriate combinations of magnetic fields, electric fields, and in-plane uniaxial strain shown in the last column of Table \ref{tab:summary_table}. Moreover, as we discussed above, if the iCDW order also breaks translational symmetry along the $c$-axis, these uniform magnetic states become ``antiferromagnetic," in the sense that they acquire modulation along the $c$-axis

Our results therefore provide concrete guidance to experimentally establish the type of loop-current order realized in \AVS{}. In this regard, we note that while experiments have reported threefold rotational symmetry breaking deep inside the charge-ordered phase of certain kagome metals~\cite{Xiang2021,Zhao2021Cascade,Li2022Rotation,Nie2022Charge,Xu_MOKE}, which seems consistent with the \Q{} state, it is important to establish whether this symmetry-breaking is a bulk or surface phenomenon, and whether it persists in the absence of applied magnetic fields, which could imply one of the \QQ{} states. Establishing the type of iCDW order realized in these compounds is important not only to identify the dominant interactions between the various vHs present near the Fermi level, but also to elucidate the properties of the superconducting state that onsets deep inside the charge-ordered phase. 

Indeed, the type of subsidiary uniform order realized in the mixed iCDW-rCDW state can have profound consequences for the pairing state. For instance, the ferromagnetic moments generated in the \QQ{} intra-orbital-iCDW state are expected to oppose pairing and thus strongly suppress conventional superconductivity. On the other hand, the orthorhombic distortion in the \Q{} state can mix pairing states of different symmetries, and even induce nodes in a chiral superconducting state~\cite{Guguchia2022Tunable}. As for the higher-order magnetic multipolar orders induced in the other \QQ{} states, little is known about their interplay with superconductivity. One interesting prospect is that the simultaneous presence of both electric and magnetic dipole moments may harbor the unusual pair-density wave (PDW) state~\cite{Agterberg2003}. Interestingly, a PDW was recently proposed to be realized in these compounds~\cite{Chen2021Roton}. More broadly, the \AVS{} systems offer a promising framework to investigate and elucidate how pairing is modified by exotic types of uniform magnetic order.

\begin{acknowledgments}
We acknowledge helpful discussions with M. I. Aroyo and M. Geier. M.H.C. has received funding from the European Union's Horizon 2020 research and innovation programme under the Marie Sklodowska-Curie grant agreement No 101024210. T.B. was supported by a NSF CAREER grant DMR-2046020. B.M.A. acknowledges support from a research grant (40509) from VILLUM FONDEN. R.M.F. was supported by the Air Force Office of Scientific Research under Award No. FA9550-21-1-0423. 
\end{acknowledgments}

\appendix

\section{First Principles Methods} \label{app:DFT}

First-principles density functional theory (DFT) calculations were performed using the Vienna ab initio Simulation Package (VASP)~\cite{VASP1, VASP2}. Further details of the first principles calculations were discussed in Ref.~\onlinecite{Christensen2021}. The band structure shown in Fig.~\ref{fig:orbitals_and_bands}(c) is that of CsV$_3$Sb$_5$. Since the effect of the alkali metal is mostly geometric in these systems, the band structure of KV$_3$Sb$_5$ and RbV$_3$Sb$_5$ are similar~\cite{Ortiz2019New}. While it is possible that small energy shifts change the order of the vHs, this does not affect the present discussion. 

The irreps of the DFT bands were obtained by using the Kohn-Sham band wavefunctions at the $M$ point, and finding their transformation properties under space group by inspection. The resultant irreps and orbital characters were confirmed to be consistent with each other by using the induced band representations approach~\cite{Bradlyn2017}.

\section{Transformation properties of possible order parameters}\label{app:space_group_irreps}

In this section, we derive the transformation properties (and hence the irreps) corresponding to the different order parameters, by starting from the individual atomic orbitals of the V atoms and then building the Hermitian bilinears. For the sake of brevity, we only focus on the two bands that give rise to the vHs below the Fermi level, but the same procedure can be extended to include the ones above the Fermi level as well. 

We consider ideal kagome layers stacked with atoms aligned on top of each other. This configuration gives rise to the space group $P6/mmm$, and corresponds to the crystal structure of \AVS{} compounds. The unit cell and the lattice vectors are shown in Fig.~\ref{fig:orbitals_and_bands}(a). There are three symmetry-equivalent V atoms in each unit cell. The $k_z=0$ plane of the Brillouin zone is shown in Fig.~\ref{fig:orbitals_and_bands}(b). There are two bands near the Fermi level at the M points, which transform as $M_1^+$ and $M_4^+$ irreps of the space group. Note that we use the convention of the Cracknell-Davies-Miller-Love tables~\cite{Cracknell1979Kronecker, Aroyo2006} available in the tables of the Bilbao Crystallographic Server, whereas another commonly used resource is the book of Koster~\cite{Koster1963}. The irreps of the little group of M in both conventions are shown in Table~\ref{table:irrep}.

\begin{table}
\begin{tabular}{C{0.24\columnwidth}C{0.24\columnwidth}|C{0.16\columnwidth}C{0.16\columnwidth}C{0.16\columnwidth}}
    CDML~\cite{Cracknell1979Kronecker} & Koster~\cite{Koster1963}  & $2_{001}$     & $2_{010}$     & $\bar{1}$\\
\hline
$M_1^\mp$ & $M_1^\mp$ & $+1$             & $+1$             & $\mp1$     \\
$M_2^\mp$ & $M_3^\mp$ & $+1$             & $-1$            & $\mp1$     \\
$M_3^\mp$ & $M_4^\mp$ & $-1$            & $+1$             & $\mp1$     \\
$M_4^\mp$ & $M_2^\mp$ & $-1$            & $-1$            & $\mp1$     \\
\end{tabular}
	\caption{Characters of the irreps of the little group of the M$_1$ point (similarly to Table \ref{tab:irrep} in the main text). We only list the characters for the three generators of the little group, and these irreps correspond to the space group irreps by the same name. Note that the space group irreps are three-dimensional, because the star of M has three distinct wavevectors. The only ambiguity upon going from the little group irreps to the space group irreps is possible minus signs in the off-diagonal elements, which are trivial up to a coordinate transformation, and do not change the characters of the irreps. Time-reversal (TR) odd irreps (not shown) are denoted by a $m$ prefix as $mM_i^\mp$.}
	\label{table:irrep}
\end{table}

The $M_1^+$ and the $M_4^+$ bands near the Fermi level at M are formed by vanadium $d_{z^2}$ and $d_{xz}$ orbitals, respectively. These orbitals transform as the $A_{1g}$ and $B_{2g}$ (or $B_{3g}$, depending on the axis choice) irreps of the $mmm$ site symmetry group of the V-site (see Fig.~\ref{fig:orbitals_and_bands}). We denote the annihilation operator for the $A_{1g}$ orbital on atom $V_i$ in the unit cell at $R=m\cdot\vec{a}+n\cdot\vec{b}$ as $a_{i\sigma,(m,n)}$, where $\sigma$ is the spin component along $\hat{z}\parallel \vec{c}$. By inspection, i.e., by considering the effect of different symmetry operations on the positions of the atoms and on the alignment of the orbitals, we find that 
\begin{equation*}
    \begin{split}
        \bar{1}a^\dagger_{1\sigma,(0,0)} \bar{1}^{-1} &= a^\dagger_{1\sigma,(0,1)} \\
        2_{001}a^\dagger_{1\sigma,(0,0)} 2_{001}^{-1} &= (\sigma i) a^\dagger_{1\sigma,(0,1)} \\
        2_{010}a^\dagger_{1\sigma,(0,0)} 2_{010}^{-1} &= (+i) a^\dagger_{1-\sigma,(0,0)}
    \end{split}
\end{equation*}
\begin{equation}
    \begin{split}
        \bar{1}a^\dagger_{2\sigma,(0,0)} \bar{1}^{-1} &= a^\dagger_{2\sigma,(-1,0)} \\
        2_{001}a^\dagger_{2\sigma,(0,0)} 2_{001}^{-1} &= (\sigma i) a^\dagger_{2\sigma,(-1,0)} \\
        2_{010}a^\dagger_{2\sigma,(0,0)} 2_{010}^{-1} &= (+i) a^\dagger_{3-\sigma,(0,0)}
    \end{split}
\end{equation}
\begin{equation*}
    \begin{split}
        \bar{1}a^\dagger_{3\sigma,(0,0)} \bar{1}^{-1} &= a^\dagger_{3\sigma,(1,1)} \\
        2_{001}a^\dagger_{3\sigma,(0,0)} 2_{001}^{-1} &= (\sigma i) a^\dagger_{3\sigma,(1,1)} \\
        2_{010}a^\dagger_{3\sigma,(0,0)} 2_{010}^{-1} &= (+i) a^\dagger_{2-\sigma,(0,0)}
    \end{split}
\end{equation*}
where the spin-1/2 nature of the electrons are taken into account so that rotations by $\pi$ bring in a phase factor of $\mp i$. Similarly, we denote the annihilation operator for the $B_{2g}$ orbital on atom V$_i$ in the unit cell at $R=m\cdot\vec{a}+n\cdot\vec{b}$ as $b_{i\sigma,(m,n)}$. By inspection, we find that
\begin{equation*}
    \begin{split}
        \bar{1}b^\dagger_{1\sigma,(0,0)} \bar{1}^{-1} &= b^\dagger_{1\sigma,(0,1)} \\
        2_{001}b^\dagger_{1\sigma,(0,0)} 2_{001}^{-1} &= (-\sigma i) b^\dagger_{1\sigma,(0,1)} \\
        2_{010}b^\dagger_{1\sigma,(0,0)} 2_{010}^{-1} &= (-i) b^\dagger_{1-\sigma,(0,0)}
    \end{split}
\end{equation*}
\begin{equation}
    \begin{split}
        \bar{1}b^\dagger_{2\sigma,(0,0)} \bar{1}^{-1} &= b^\dagger_{2\sigma,(-1,0)} \\
        2_{001}b^\dagger_{2\sigma,(0,0)} 2_{001}^{-1} &= (-\sigma i) b^\dagger_{2\sigma,(-1,0)} \\
        2_{010}b^\dagger_{2\sigma,(0,0)} 2_{010}^{-1} &= (+i) b^\dagger_{3-\sigma,(0,0)}
    \end{split}
\end{equation}
\begin{equation*}
    \begin{split}
        \bar{1}b^\dagger_{3\sigma,0} \bar{1}^{-1} &= b^\dagger_{3\sigma,(1,1)} \\
        2_{001}b^\dagger_{3\sigma,0} 2_{001}^{-1} &= (-\sigma i) b^\dagger_{3\sigma,(1,1)} \\
        2_{010}b^\dagger_{3\sigma,0} 2_{010}^{-1} &= (+i) b^\dagger_{2-\sigma,(0,0)}
    \end{split}
\end{equation*}
Note that $a^\dagger$ and $b^\dagger$ are creation operators for atomic orbitals localized in real space. The momentum-space counterparts that create electrons on Bloch states with wavevector $\vec{q}=q_1 \vec{a}^* + q_2 \vec{b}^*$ are defined as  
\begin{equation}
	c_{i\sigma \vec{q}}^\dagger = \sum_{m,n} a_{i\sigma (m,n)}^\dagger \exp \left\{ -i (q_1m+q_2n)\right\}
\end{equation}
\begin{equation}
        d_{i\sigma \vec{q}}^\dagger = \sum_{m,n} b_{i\sigma (m,n)}^\dagger \exp \left\{ -i (q_1m+q_2n)\right\}
\end{equation}
We focus solely on the $\mbf{k}=\mbf{Q}_{1}=\mbf{a^*}/2$ point in the Brillouin zone, and drop the $\vec{q}$ subscript in the $c$ and $d$ operators. The transformation properties of these momentum-space operators are
\begin{equation*}
    \begin{split}
        \bar{1}c^\dagger_{1\sigma} \bar{1}^{-1} &= c^\dagger_{1\sigma} \\
        2_{001}c^\dagger_{1\sigma} 2_{001}^{-1} &= (\sigma i) c^\dagger_{1\sigma} \\
        2_{010}c^\dagger_{1\sigma} 2_{010}^{-1} &= (+i) c^\dagger_{1-\sigma}
    \end{split}
\end{equation*}
\begin{equation}
    \begin{split}
        \bar{1}c^\dagger_{2\sigma} \bar{1}^{-1} &= -c^\dagger_{2\sigma} \\
        2_{001}c^\dagger_{2\sigma} 2_{001}^{-1} &= (-\sigma i) c^\dagger_{2\sigma} \\
        2_{010}c^\dagger_{2\sigma} 2_{010}^{-1} &= (+i) c^\dagger_{3,-\sigma}
\end{split}
\end{equation}
\begin{equation*}
    \begin{split}
        \bar{1}c^\dagger_{3\sigma} \bar{1}^{-1} &= -c^\dagger_{3\sigma} \\
        2_{001}c^\dagger_{3\sigma} 2_{001}^{-1} &= (-\sigma i) c^\dagger_{3\sigma} \\
        2_{010}c^\dagger_{3\sigma} 2_{010}^{-1} &= (+i) c^\dagger_{2-\sigma}
    \end{split}
\end{equation*}
as well as
\begin{equation*}
    \begin{split}
        \bar{1}d^\dagger_{1\sigma} \bar{1}^{-1} &= d^\dagger_{1\sigma} \\ 
        2_{001}d^\dagger_{1\sigma} 2_{001}^{-1} &= (-\sigma i) d^\dagger_{1\sigma} \\
        2_{010}d^\dagger_{1\sigma} 2_{010}^{-1} &= (-i) d^\dagger_{1-\sigma}
    \end{split}
\end{equation*}
\begin{equation}
    \begin{split}
        \bar{1}d^\dagger_{2\sigma} \bar{1}^{-1} &= -d^\dagger_{2\sigma} \\
        2_{001}d^\dagger_{2\sigma} 2_{001}^{-1} &= (\sigma i) d^\dagger_{2\sigma} \\
        2_{010}d^\dagger_{2\sigma} 2_{010}^{-1} &= (+i) d^\dagger_{3-\sigma}
    \end{split}
\end{equation}
\begin{equation*}
    \begin{split}
        \bar{1}d^\dagger_{3\sigma} \bar{1}^{-1} &= -d^\dagger_{3\sigma} \\
        2_{001}d^\dagger_{3\sigma} 2_{001}^{-1} &= (\sigma i) d^\dagger_{3\sigma} \\
        2_{010}d^\dagger_{3\sigma} 2_{010}^{-1} &= (+i) d^\dagger_{2-\sigma}
    \end{split}
\end{equation*}
The transformation properties of the bilinears follow from these equations.
\newpage

\bibliography{rCDW_iCDW}

\end{document}